\documentclass{appolb}
\usepackage{epsf,rotate}
\usepackage{amsmath,amstext,amsbsy,amsopn,amsfonts,amssymb}
\usepackage[dvips]{graphicx}
\usepackage[latin1]{inputenc}
\usepackage[english]{babel}
\usepackage[T1]{fontenc}
\usepackage{psfrag}

\newcommand{\Log}{\mbox{Log}}
\newcommand{\RRe}{\mbox{Re}}
\newcommand{\IIm}{\mbox{Im}}

\newcommand{\eqdef}{\mbox{$ \stackrel{\rm def}{=}$}}
\newcommand{\asbar}{\ifmmode\wbar {\alpha}_{\rm s}\else{$\wbar{\alpha}_{\rm s}$}\fi}

 \catcode`\@=11

\def \lab #1 {\label{#1}}
\newcommand \ci [1] {\cite{#1}}

\def \Im {\mathop{\rm Im}\nolimits}
\def \Re {\mathop{\rm Re}\nolimits}

\def \qqquad {\qquad\quad}
\def \e  {\mathop{\rm e}\nolimits}
\newcommand\lr[1]{{\left({#1}\right)}}
\newcommand \wbar [1] {\overline{#1}}
\newcommand \vev [1] {\langle{#1}\rangle}

\newcommand \ket [1] {|{#1}\rangle}

\newcommand \mybf[1] {\mbox{\boldmath$ {#1} $}}

\newcommand \ot[1] {\Hat{t}_{#1}}
\newcommand \otb[1] {\Hat{\wbar{t}}_{#1}}
\newcommand \oq[1] {\Hat{q}_{#1}}
\newcommand \oqb[1] {\Hat{\wbar{q}}_{#1}}

\def \SL  {\mbox{SL}}

\def\numberbysection{\@addtoreset{equation}{section}
                     \def\theequation{\thesection\arabic{equation}}}

\begin{document}
\eqsec
\title{Classification of four-Reggeon states in multi-colour QCD
\author{{\bf Jan Kota{\'n}ski}
\address{M. Smoluchowski Institute of Physics, Jagellonian University\\
Reymonta 4, 30-059 Krak{\'o}w, Poland}}}


\maketitle
                                                                  
\begin{abstract}
{\normalsize
$N-$Reggeized gluon states in Quantum Chromodynamics are described by BKP
equation. In order to solve this equation for $N>3$ particles
the $Q-$Baxter operator method is used. Spectrum
of the integrals of motion of the system exhibits
a complicated structure.
In this work we consider the case with $N=4$ Reggeons
where complicated relations between $q_3-$spectrum and $q_4-$spectrum
are analysed. Moreover, corrections to WKB approximation for 
$N=4$ and $q_3=0$ are computed.}
\end{abstract}

\newline

\vspace*{1cm}
\noindent TPJU-04/2006 \newline

\newpage

\bibliographystyle{unsrt}

\section{Introduction}

In the Regge limit, where the total energy $s$ is large while
the transfer of four-momentum $t$ is low and fixed,
the scattering amplitude of two hadrons can be rewritten
as an exchange of effective particles, \ie propagating in 
$t-$channel reggeized gluons, which are also called Reggeons 
\ci{gell,gell2,Gribov:1968fc,Fadin:1975cb,Bartels:1977hz,Cheng:1977gt}.

Due to ordering of generalized leading logarithm approximation 
contribution from exchange  of $N-$Reggeon states is suppressed
by factor $\alpha_s^{N-2}$ where $\alpha_s$ is a strong coupling
constant.
Thus, the leading contribution is given by two Reggeon states,
\ie BFKL Pomeron. The equation describing this case was
firstly derived and solved by
Balitsky, Fadin, Kuraev and Lipatov 
\ci{Balitsky:1978ic,Kuraev:1977fs,Fadin:1975cb}.
Extension of this equation for more than two Reggeons
was formulated in 1980  
\ci{Bartels:1980pe,Kwiecinski:1980wb,Jaroszewicz:1980rw}
by Bartels, Kwieci\'nski, Prasza{\l}owcz and Jaroszewicz.
It has a structure of Schr\"odinger equation.
The first solutions for exchange of three Reggeons 
appear in nineties 
\ci{Janik:1998xj,Gauron:1987jt,Lukaszuk:1973nt}
and these solutions correspond to leading
contribution to the QCD odderon state
as well as subleading contribution to the Pomeron state.

The Schr\"odinger equation for $N\ge4$ Reggeons
contains complicated colour factor and even
in 't Hooft's multi-colour limit
\ci{'tHooft:1973jz,Lipatov:1990zb,Lipatov:1993qn},
\ie $N_c \to \infty$,
formulating of its solutions poses
real 
challenge and demands using of
advanced methods of integrable systems.
In the multi-colour limit
the Reggeon Hamiltonian corresponds
to the Hamiltonian $\SL(2,\mathbb{C})$
spin magnet which turns out to be solvable model.
Thus, making use of $Q-$Baxter and
Separation of Variable methods
one can solve the multi-Reggeon problem
is multi-colour limit.
The solutions for higher $N=4,\ldots,8$
were obtained
recently in series of papers 
\ci{Derkachov:2002pb,Korchemsky:2001nx,Derkachov:2002wz} 
written in collaboration with S.\'E. Derkachov,
G.P. Korchemsky and A.N. Manashov.
Similar results for $N=4$ appear also in 
Ref.~\ci{deVega:2002im,DeVega:2001pu}.

The present work is continuation of 
\ci{Kotanski:2006ec}
as well as
\ci{Derkachov:2002pb,Korchemsky:2001nx,Derkachov:2002wz}.
We concentrate on $N=4$ case
with conformal Lorentz spin $n_h=0$.
Here, classification of the four Reggeon states 
is performed and  corrections to 
the WKB calculation are computed.
Moreover, we calculate
the rich spectrum of the the Reggeon energy
and the conformal charges $\{q_3, q_4\}$ for four reggeized gluons.
For a broader perspective see also Ref.~\ci{Kotanski:2005ci}.

Thus, in Section 2 we explain when the reggeization of the gluon appears 
\ci{Braun:1994ll}.
Next, we perform the multi-colour limit 
\ci{'tHooft:1973jz,Lipatov:1990zb,Lipatov:1993qn}, 
discuss the properties of the 
$\SL(2,\mathbb{C})$ symmetry and construct invariants of this symmetry,
the conformal charges. In Section 3 we introduce 
a Baxter
$Q-$operator method \ci{Baxter} with
Baxter equations that allow us to solve the Reggeon system, completely.
Next, we present the exact solution to the Baxter equations.
It consists in rewriting the Baxter equation into the differential
equation, which may be solved by a series method. We find
the quantization conditions for $\{q_3, q_4\}$ 
which come analyticity properties of the Baxter functions.
We also recapitulate known properties of $N-$Reggeon spectrum.
The numerical results 
are shown
in Section 6 \ci{Korchemsky:2001nx,Derkachov:2002wz}.
In particular, we discuss the resemblant and winding structures
of the $q_4$ and $q_3$ spectrum and 
also  corrections to 
the WKB approximation for $q_3=0$.
At the end we make final conclusions.

\section{System with $\SL(2,\mathbb{C})$ symmetry}

\subsection{Hamiltonian}
In the Regge limit 
\begin{equation}
s \to \infty \mbox{ and } t=\mbox{const}
\lab{eq:Rlim}
\end{equation}
$N-$Reggeon Hamiltonian can be rewritten
as a sum of BFKL kernels.
Performing the multi-colour limit 
\ci{'tHooft:1973jz,Lipatov:1990zb,Lipatov:1993qn},
where a number of colours $N_c \to \infty$,
the system of $N-$reggeized gluons is described by Hamiltonian
\begin{equation}
\mathcal{H}_{N}=H_{N}+\wbar{H}_{N}\mbox {,}\qquad 
[H_{N},\wbar{H}_{N}]=0\mbox { }
\lab{eq:sepH}
\end{equation}
which can be written
in terms of the conformal spins (\ref{eq:spins}):
\begin{equation}
H_{N}=\sum _{k=1}^{N}H(J_{k,k+1})\mbox {,}\quad 
\wbar{H}_{N}=\sum _{k=1}^{N}H(\wbar{J}_{k,k+1})\mbox {,}
\lab{eq:Ham}
\end{equation}
where
\begin{equation}
H(J)=\psi (1-J)+\psi (J)-2\psi (1)
\end{equation}
with $\psi (x)=d\ln \Gamma (x)/dx$ being the Euler function and
$J_{N,N+1}=J_{N,1}$. Here operators, $J_{k,k+1}$ and $\wbar{J}_{k,k+1}$,
are defined through the Casimir operators for the sum of the spins
of
the neighbouring Reggeons
\begin{equation}
J_{k,k+1}(J_{k,k+1}-1)=(S^{(k)}+S^{(k+1)})^{2}
\lab{eq:jvss}
\end{equation}
with $S_{\alpha }^{(N+1)}=S_{\alpha }^{(1)}$, and $\wbar{J}_{k,k+1}$ is
defined similarly.

\subsection{Symmetry $\SL(2,\mathbb{C})$}

The Hamiltonian (\ref{eq:sepH}) is invariant under 
the coordinate transformation of the $\SL(2,\mathbb{C})$ group
\begin{equation}
z_{k}^{\prime }=\frac{az_{k}+b}{cz_{k}+d}\mbox {,}\qquad 
\wbar{z}_{k}^{\prime }=
\frac{\wbar{a}\wbar{z}_{k}+\wbar{b}}{\wbar{c}\wbar{z}_{k}+\wbar{d}}
\lab{eq:trcoords}
\end{equation}
with $k=1,\ldots ,N$ and  $ad-bc=\wbar{a}\wbar{d}-\wbar{b}\wbar{c}=1$.

Now, one may associate with the $k-$th particle 
the generators of this transformation \ci{Derkachov:2001yn}.
These generators are a pair of mutually commuting holomorphic
and anti-holomorphic spin operators, 
$S_{\alpha }^{(k)}$ and $\wbar{S}_{\alpha }^{(k)}$.
They satisfy the standard commutation relations 
$\left[S_{\alpha }^{(k)},S_{\beta }^{(n)}\right]
=i\epsilon _{\alpha \beta \gamma }\delta ^{kn}S_{\gamma }^{(k)}$
and similarly for $\wbar{S}_{\alpha }^{(k)}$.
The generators act on the quantum
space of the $k$-th particle, $V^{(s_{k},\wbar{s}_{k})}$ as 
differential operators 
\begin{equation}
\begin{array}{ccc}
 S_{0}^{k}=z_{k}\partial _{z_{k}}+s_{k}\,, \quad  & 
S_{-}^{(k)}=-\partial _{z_{k}}\,, \quad  & 
S_{+}^{(k)}=z_{k}^{2}\partial _{z_{k}}+2s_{k}z_{k}\,,\\
 \wbar{S}_{0}^{k}=
\wbar{z}_{k}\partial _{\wbar{z}_{k}}+\wbar{s}_{k}\,, \quad  & 
\wbar{S}_{-}^{(k)}=-\partial _{\wbar{z}_{k}} \,, \quad  & 
\wbar{S}_{+}^{(k)}=
\wbar{z}_{k}^{2}\partial _{\wbar{z}_{k}}+2\wbar{s}_{k}\wbar{z}_{k}\,,\\
\end{array}
\lab{eq:spins}
\end{equation}
 where $ S_{\pm}^{(k)}=S_{1}^{(k)}\pm i S_{2}^{(k)}$
while the complex parameters, $s_{k}$ and $\wbar{s}_{k}$,
are called the complex spins. Thus, the Casimir operator reads 
\begin{equation}
\sum_{j=0}^2(S_j^{(k)})^{2}=(S_{0}^{(k)})^{2}+(S_{+}^{(k)}S_{-}^{(k)}
+S_{-}^{(k)}S_{+}^{(k)})/2=s_{k}(s_{k}-1)
\lab{eq:casimir}
\end{equation}
and similarly for the anti-holomorphic operator $(\wbar{S}^{(k)})^{2}$.

The eigenstates of $\SL(2,\mathbb{C})$ invariant system transform
as \ci{CFT,Zuber:1995rj}
\begin{equation}
\Psi (z_{k},\wbar{z}_{k})\rightarrow \Psi ^{\prime }(z_{k},\wbar{z}_{k})
=(cz_{k}+d)^{-2s_{k}}(\wbar{c}\wbar{z}_{k}
+\wbar{d})^{-2\wbar{s}_{k}}\Psi 
(z_{k}^{\prime },
\wbar{z}_{k}^{\prime })\mbox {.}
\lab{eq:trpsi}
\end{equation}

In statistical physics  (\ref{eq:Ham}) is called the Hamiltonian of 
the non-compact $\SL(2,\mathbb{C})$
XXX Heisenberg spin magnets.
It describes the  nearest neighbour interaction
between $N$ non-compact $\SL(2,\mathbb{C})$ spins attached to the particles
with periodic boundary conditions.

For the homogeneous spin chain we have to take $s_{k}=s$ and 
$\wbar{s}_{k}=\wbar{s}$. In QCD values of $(s,\wbar{s})$ depend on 
a chosen scalar product in the space of the wave-functions (\ref{eq:trpsi}) 
and they are usually equal to $(0,1)$ or $(0,0)$ 
\ci{Derkachov:2001yn,DeVega:2001pu}.

\subsection{Scalar product}

In order to find the high energy behaviour of the scattering amplitude 
we have to solve the Schr\"{o}dinger equation 
\begin{equation}
\mathcal{H}^{(s=0,\wbar s=1)}_{N}
\Psi (\vec{z}_{1},\vec{z}_{2},\ldots ,\vec{z}_{N})
=E_{N}\Psi (\vec{z}_{1},\vec{z}_{2},\ldots ,\vec{z}_{N})
\lab{eq:Schr}
\end{equation}
with the eigenstate $\Psi (\vec{z}_{1},\vec{z}_{2},\ldots ,\vec{z}_{N})$
being a single-valued function on the plane $\vec{z}=(z,\wbar{z})$,
normalizable with respect to the $\SL(2,\mathbb{C})$ invariant scalar
product\begin{equation}
||\Psi ||^{2}=\vev{\Psi|\Psi}=
\int d^{2}z_{1}d^{2}z_{2}\ldots d^{2}z_{N}
|\Psi (\vec{z}_{1},\vec{z}_{2},\ldots ,\vec{z}_{N})|^{2}\,,
\lab{eq:norm}
\end{equation}
where $d^{2}z_i=dx_idy_i=dz_id\wbar{z}_i/2$ with $\wbar z_i={z_i}^{\ast}$.
One may notice that it is 
possible to use other scalar products corresponding  to different choice
of $(s,\wbar s)$.
For farther information see Refs.~\ci{Derkachov:2002wz,Kotanski:2005ci}
and \ci{deVega:2002im,DeVega:2001pu}.

Let us consider the amplitude for the scattering of two colourless objects $A$
and $B$. In the Regge limit, the contribution
to the scattering amplitude from $N-$gluon exchange in the $t-$channel takes
the form 
\begin{equation}
\mathcal{A}(s,t)=is \sum_N (i {\alpha}_s)^N \mathcal{A}_N(s,t)\,.
\lab{eq:a-st}
\end{equation}
Using the $\SL(2,\mathbb{C})$ scalar product (\ref{eq:norm}) we have
\begin{equation}
\mathcal{A}_N(s,t) = s \int d^2 z_0\,
\e^{i\vec z_0 \cdot \vec p}
\vev{\tilde{\Phi}_A(\vec z_0)|
\e^{- \wbar{\alpha}_s Y {\cal H}_N/4 }
 | \tilde{\Phi}_B(0)}\,,
\lab{A}
\end{equation}
where the rapidity $Y=\ln s$.
Here the Hamiltonian ${\cal H}_N$ is 
related to the sum of $N$ BFKL kernels
corresponding to nearest neighbour interaction between $N$ reggeized gluons.
The wave-functions 
$\ket{\Phi_{A(B)}(\vec z_0)}\equiv \Phi_{A(B)}(\vec z_i-\vec z_0)$
describe the coupling of $N-$ gluons to the scattered particles.
The $\vec z_0 -$ integration fixes the momentum transfer $t=-\vec p^{\,2}$.

\subsection{Conformal charges $q_k$ and the conformal spins}
The Hamiltonian (\ref{eq:sepH}) possesses 
a complete set of the integrals of motion 
$\{\vec p, \vec q_k\}$ where 
$\vec{q}_k=\{q_k,\wbar{q}_k \}$ with  $k=2,\ldots N$ are called 
conformal charges while $\vec p=\{p,\wbar p\}$ is 
the total momentum of the system. 
In order to construct them we introduce the Lax operators 
\ci{Sklyanin:1991ss,Faddeev:1994nk,Faddeev:1996iy,Faddeev:1979gh} 
in holomorphic and anti-holomorphic sectors: 
\begin{eqnarray}
 L_{k}(u)&=&u+i(\sigma \cdot S^{(k)})=\left(\begin{array}{cc}
 u+iS_{0}^{(k)} & iS_{-}^{(k)}\\
 iS_{+}^{(k)} & u-iS_{0}^{(k)}\end{array}\right)\,,
\nonumber\\
 \wbar{L}_{k}(\wbar{u})&=&
\wbar{u}+i(\sigma \cdot \wbar{S}^{(k)})=\left(\begin{array}{cc}
 \wbar{u}+i\wbar{S}_{0}^{(k)} & i\wbar{S}_{-}^{(k)}\\
 i\wbar{S}_{+}^{(k)} & \wbar{u}-i\wbar{S}_{0}^{(k)}
\end{array}\right)
\lab{eq:Lax}
\end{eqnarray}
 with $u$ and $\wbar{u}$ being arbitrary complex parameters  called 
the spectral parameters and 
$\sigma _{\alpha }$ being
Pauli matrices.

To identify the total set of the integrals of motion of the model,
one constructs the auxiliary holomorphic monodromy matrix
\begin{equation}
T_{N}(u)=L_{1}(u)L_{2}(u)\ldots L_{N}(u)
\lab{eq:LLL}
\end{equation}
and similarly for, the anti-holomorphic monodromy 
operator $\wbar{T}_{N}(\wbar{u})$.
Taking the trace of the monodromy matrix we define the auxiliary transfer
matrix (spectral invariants)
\begin{equation}
\ot{N}(u)
=\mbox {tr}\left[T_{N}(u)\right]=2u^{N}+\oq{2} u^{N-2}+ \ldots +\oq{N}
\lab{eq:tnu}
\end{equation}
and similarly for $\otb{N}(\wbar{u})$. 
We see from (\ref{eq:tnu})
the advantage of using the transfer matrix $\ot{N}(u)$:
that is 
a polynomial in $u$ with coefficients given in terms of conformal charges
$\oq{k}$ and $\oqb{k}$, 
which are expressed as linear
combinations of the products of $k$ spin operators:
\begin{eqnarray}
\displaystyle
\nonumber
\oq{2} & = & 
-2\sum _{i_{2}>i_{1}=1}^{N}\sum _{j_{1}=0}^{2}
\left(S_{j_{1}}^{(i_{1})}S_{j_{1}}^{(i_{2})}\right)\\
\displaystyle
\nonumber
\oq{4} & = & 
-\sum _{i_{2}>i_{1}=1}^{N} \sum _{i_{4}>i_{3}=1}^{N}\sum _{j_{1},j_{2}=0}^{2}
\varepsilon _{i_{1}i_{2}i_{3}i_{4}}
\left(S_{j_{1}}^{(i_{1})}S_{j_{1}}^{(i_{2})}\right)
\left(S_{j_{2}}^{(i_{3})}S_{j_{2}}^{(i_{4})}\right)\\
\displaystyle
\nonumber
\oq{6} & = & 
-\frac{1}{3}\sum _{i_{2}>i_{1}=1}^{N}\sum _{i_{4}>i_{3}=1}^{N}
\sum _{i_{6}>i_{5}=1}^{N}\sum _{j_{1},j_{2},j_{3}=0}^{2}
\varepsilon _{i_{1}i_{2}i_{3}i_{4}i_{5}i_{6}}
\left(S_{j_{1}}^{(i_{1})}S_{j_{1}}^{(i_{2})}\right)\\
& & \times
\left(S_{j_{2}}^{(i_{3})}S_{j_{2}}^{(i_{4})}\right)
\left(S_{j_{3}}^{(i_{5})}S_{j_{3}}^{(i_{6})}\right)\,,
\lab{eq:q2q4q6}
\end{eqnarray}
 where $\varepsilon _{i_{1}i_{2}\ldots i_{k}}$ 
is completely anti-symmetric
tensor and $\varepsilon _{i_{1}i_{2}\ldots i_{k}}=1$ 
for $i_{1}<i_{2}<\ldots <i_{k}$.
So for even $k$ we have a formula for conformal charges 
\begin{align}
\oq{k} & =-\frac{2}{(k/2)!}
\sum_{
\begin{array}{c}
 i_{2}>i_{1}=1\\
 i_{4}>i_{3}=1\\
 \ldots \\
 i_{n}>i_{n-1}=1
\end{array}}^{N}
\sum _{j_{1},j_{2},\ldots ,j_{k/2}=0}^{2}
\varepsilon _{i_{1}i_{2}\ldots i_{k}}
\prod_{n=1}^{k/2}
\left(S_{j_{n}}^{(i_{2 n-1})}S_{j_{n}}^{(i_{2 n})} \right)\,.
\lab{eq:sqen}
\end{align}

For odd $k$'s we have
\begin{eqnarray}
\nonumber
\oq{3}&=&2\sum _{i_{1},i_{2},i_{3}=1}^{N}
\varepsilon _{i_{1}i_{2}i_{3}}
\left(S_{0}^{(i_{1})}S_{1}^{(i_{2})}S_{2}^{(i_{3})}\right)\\
\nonumber
& = &
2\sum _{i_{3}>i_{2}>i_{1}=1}^{N}
\sum _{j_{1},j_{2},j_{3}=0}^{2}
\varepsilon _{i_{1}i_{2}i_{3}}
\varepsilon _{j_{1}j_{2}j_{3}}
\left(S_{j_{1}}^{(i_{1})}S_{j_{2}}^{(i_{2})}S_{j_{3}}^{(i_{3})}\right)\,,\\
\nonumber
\oq{5} & = & -2\sum _{i_{3},i_{2},i_{1}=1}^{N}
\sum _{i_{5}>i_{4}=1}^{N}\sum _{j_{4}=0}^{2}
\varepsilon _{i_{1}i_{2}i_{3}i_{4}i_{5}}
\left(S_{0}^{(i_{1})}S_{1}^{(i_{2})}S_{2}^{(i_{3})}\right)
\left(S_{j_{4}}^{(i_{4})}S_{j_{4}}^{(i_{5})}\right)\,,\\
\nonumber
\oq{7} & = & \sum _{i_{3},i_{2},i_{1}=1}^{N}
\sum _{i_{5}>i_{4}=1}^{N}\sum _{i_{7}>i_{6}=1}^{N}
\sum _{j_{4},j_{5}=0}^{2}
\varepsilon _{i_{1}i_{2}i_{3}i_{4}i_{5}i_{6}i_{7}}
\left(S_{0}^{(i_{1})}S_{1}^{(i_{2})}S_{2}^{(i_{3})}\right) \\
& & \times
\left(S_{j_{4}}^{(i_{4})}S_{j_{4}}^{(i_{5})}\right)
\left(S_{j_{5}}^{(i_{6})}S_{j_{5}}^{(i_{7})}\right)
\lab{eq:q3q5q7}
\end{eqnarray}
and the general expression for an odd number of the conformal spins is 
\begin{eqnarray}
\nonumber
\oq{k} & = & \frac{2 (-1)^{(k+1)/2}}{\left(\frac{k-3}{2}\right)!}
\sum _{i_{3},i_{2},i_{1}=1}^{N}\sum _{\begin{array}{c}
 i_{5}>i_{4}=1\\
 i_{7}>i_{6}=1\\
 \ldots \\
 i_{k}>i_{k-1}=1\end{array}}^{N}\sum _{j_{1},j_{2},
\ldots ,j_{(k-3)/2}=0}^{2}
\varepsilon _{i_{1}i_{2}\ldots i_{k}}\\
 &  & \times \, 
\left(S_{0}^{(i_{1})} S_{1}^{(i_{2})} S_{2}^{(i_{3})}\right)
\prod_{n=1}^{(k-3)/2}
\left(S_{j_{n}}^{(i_{2n+2})}S_{j_{n}}^{(i_{2n+3})}\right)\,.
\lab{eq:sqok}
\end{eqnarray}

In the above formulae we have two basic blocks 
$\left(S_{0}^{(i_{1})}S_{1}^{(i_{2})}S_{2}^{(i_{3})}\right)$ 
and $\left(S_{j_{1}}^{(i_{1})}S_{j_{1}}^{(i_{2})}\right)$
whose products are summed with antisymmetric tensor 
$\varepsilon _{i_{1}i_{2}\ldots i_{k}}$.

\subsection{Two-dimensional Lorentz spin and the scaling dimension}

The Hamiltonian (\ref{eq:Ham})
is a function of two-particle Casimir operators \ci{Derkachov:2001yn}, and
therefore, it commutes with the operators of the total spin 
$S_{\alpha }=\sum _{k}S_{\alpha }^{(k)}$
and $\wbar{S}_{\alpha }=\sum _{k}\wbar{S}_{\alpha }^{(k)}$, acting
on the quantum space of the system 
$V_{N}\equiv V^{(s_{1},\wbar{s}_{1})}\otimes V^{(s_{2},\wbar{s}_{2})}\otimes 
\ldots \otimes V^{(s_{N},\wbar{s}_{N})}$.
This implies that the eigenstates can be classified according to the
irreducible representations of the $\SL(2,\mathbb{C})$ group, 
$V^{(h,\wbar{h})}$,
parameterized by spins $(h,\wbar{h})$ \ci{Derkachov:2001yn}. 

The Hamiltonian depends on differences of particle coordinates 
so the eigenfunctions can be written as 
\begin{equation}
\Psi _{\vec{p}}(\vec{z}_{1},\vec{z}_{2},\ldots ,\vec{z}_{N})=
\int d^{2}z_{0}e^{i\vec{z}_{0}\cdot \vec{p}}
\Psi (\vec{z}_{1}-\vec{z}_{0},\vec{z}_{2}-\vec{z}_{0},\ldots ,
\vec{z}_{N}-\vec{z}_{0})\,.
\lab{eq:Psip}
\end{equation}
The eigenstates $\Psi (\vec{z}_{1},\vec{z}_{2},\ldots ,\vec{z}_{N})$
belonging to $V^{(h,\wbar{h})}$ are labelled by the centre-of-mass coordinate
$\vec{z}_{0}$ and can be chosen to have the following  the $\SL(2,\mathbb{C})$
transformation properties
\begin{multline}
\Psi (\{\vec{z}_{k}^{\ \prime }-\vec{z}_{0}^{\ \prime }\})=\\
=(cz_{0}+d)^{2h}(\wbar{c}\wbar{z}_{0}+\wbar{d})^{2\wbar{h}}
\left(\prod _{k=1}^{N}(cz_{k}+d)^{2s_{k}}(\wbar{c}\wbar{z}_{k}
+\wbar{d})^{2\wbar{s}_{k}}
\right)
\Psi (\{\vec{z}_{k}-\vec{z}_{0}\})
\lab{eq:trpsiz0}
\end{multline}
 with $z_{0}$ and $\wbar{z}_{0}$ transforming in the same way as $z_{k}$
and $\wbar{z}_{k}$, (\ref{eq:trcoords}).
As a consequence, they diagonalize the Casimir operators:
\begin{equation}
(S^2-h(h-1)) 
\Psi(\vec{z}_{1},\vec{z}_{2},\ldots ,\vec{z}_{N})=0
\lab{eq:SCas}
\end{equation}
corresponding to the total spin of the system, 
\begin{equation}
S^2= 
\sum _{i_{2},i_{1}=1}^{N}
\sum _{j=0}^{2}S_{j}^{(i_{1})}S_{j}^{(i_{2})}=
-\oq{2}-\sum _{k=1}^{N}s_{k}(s_{k}-1)\,.
\lab{eq:S2}
\end{equation}

The complex parameters $(s_{k},\wbar{s}_{k})$ and $(h,\wbar{h})$ parameterize
the irreducible representations of the $\SL(2,\mathbb{C})$ group. 
For the principal series representation
they satisfy the conditions 
\begin{equation}
s_{k}-\wbar{s}_{k}=n_{s_{k}}\mbox {,}\qquad s_{k}+(\wbar{s}_{k})^{*}=1
\lab{eq:ssbar}
\end{equation}
 and have the following form
\begin{equation}
s_{k}=\frac{1+n_{s_{k}}}{2}+i\nu _{s_{k}}\mbox {,}\qquad \wbar{s}_{k}=
\frac{1-n_{s_{k}}}{2}+i\nu _{s_{k}}
\lab{eq:spar}
\end{equation}
 with $\nu _{s_{k}}$being real and $n_{s_{k}}$ being integer or
half-integer. The spins $(h,\wbar{h})$ are given by similar expressions
with $n_{s_{k}}$and $\nu _{s_{k}}$ replaced by $n_{h}$ and $\nu _{h}$,
respectively
\begin{equation}
h=\frac{1+n_{h}}{2}+i\nu _{h}\mbox {,}\qquad \wbar h=
\frac{1-n_{h}}{2}+i\nu _{h}.
\lab{eq:hpar}
\end{equation}
 The parameter $n_{s_{k}}$ has the meaning of the two-dimensional
Lorentz spin of the particle, whereas $\nu _{s_{k}}$ defines its
scaling dimension. To see this one can perform a $2\pi $-rotation of
the particle on the plane, 
and find from eigenstates transformations 
(\ref{eq:trpsiz0}) that the wave-function acquires a phase. Indeed
\begin{equation}
z_{k}\rightarrow z_{k}\e^{2\pi i} 
\; 
\mbox{and} 
\;
\wbar{z}_{k}\rightarrow \wbar{z}_{k}\e^{-2\pi i}
\quad
\mbox{gives}
\quad
\Psi (z_{k},\wbar{z}_{k})\rightarrow 
(-1)^{2 n_{s_{k}}}\Psi (z_{k},\wbar{z}_{k}).
\end{equation}
For half-integer $n_{s_{k}}$ it changes the sign and the corresponding
representation is spinorial. Similarly, to define scaling dimension,
$s+\wbar{s}=1+2i\nu _{s_{k}}$
one performs the transformation 
\begin{equation}
z\rightarrow \lambda z
\quad
\mbox{and} 
\quad
\wbar{z}\rightarrow \lambda \wbar{z}
\qquad
\mbox{giving} 
\qquad
\Psi (z_{k},\wbar{z}_{k})\rightarrow \lambda ^{1+2i\nu _{s_{k}}}
\Psi (z_{k},\wbar{z}_{k})\,.
\end{equation}
Because the scalar product for the wave-functions is 
invariant under $\SL(2,\mathbb{C})$ transformations, 
(\ref{eq:trcoords}), 
the parameter $\nu _{s_{k}}$ is real.

We notice that the holomorphic and anti-holomorphic spin generators
as well as Casimir operators (\ref{eq:jvss}) are conjugated to each
other with respect to the scalar product (\ref{eq:norm}):
\begin{equation}
[S_{\alpha }^{(k)}]^{\dagger }
=-\wbar{S}_{\alpha }^{(k)}\mbox {,}\qquad [J_{k}]^{\dagger }
=1-\wbar{J}_{k}\,.
\lab{eq:sjdag}
\end{equation}
Moreover, because of the transformation law (\ref{eq:sjdag}),
$h^{\ast}=1-\wbar{h}$
\footnote{$ ^\ast$ -- denotes complex conjugation}.
This implies that $H_{N}^{\dagger }=\wbar{H}_{N}$ and, as a consequence,
the Hamiltonian is hermitian on the space of the functions endowed
with the $\SL(2,\mathbb{C})$ scalar product, 
$\mathcal{H}_{N}^{\dagger }=\mathcal{H}_{N}$.

\subsection{Conformal charges $\oq{k}$ as a differential operators}

We noticed in the previous Section 
that the conformal charge operators $\oq{k}$
are given by invariant sum of linear combinations of the products of
$k$ spin operators. They can be rewritten as $k$-th order differential
operators acting on (anti)holomorphic coordinates $(z,\wbar{z})$.

Two particle spin square can be written as
\begin{multline}
\sum _{j_{1}=0}^{2}S_{j_{1}}^{(i_{1})}S_{j_{1}}^{(i_{2})} =\\
= 
-\frac{1}{2}(z_{i_{1}}-z_{i_{2}})^{2}
\partial _{z_{i_{1}}}\partial _{z_{i_{2}}}
+(z_{i_{1}}-z_{i_{2}})
(s_{i_{2}}\partial _{z_{i_{1}}}+s_{i_{1}}\partial _{z_{i_{2}}})
+s_{1}s_{2}\,.
\end{multline}
For homogeneous spins $s=s_{1}=s_{2}=\ldots =s_{N}$,
what is also the QCD case,
we have
\begin{eqnarray}
\displaystyle
\nonumber
\oq{2}&=&-2\sum _{i_{2}>i_{1}=1}^{N}
\left(\sum _{j_{1}=0}^{2}S_{j_{1}}^{(i_{1})}S_{j_{1}}^{(i_{2})}\right)\\
\nonumber
&=&\sum _{i_{2}>i_{1}=1}^{N}\left((z_{i_{2}i_{1}})^{2(1-s)}
\partial _{z_{i_{2}}}\partial _{z_{i_{1}}}
(z_{i_{2}i_{1}})^{2s}+2s(s-1)\right)\,,
\\
\nonumber
\displaystyle
\oq{3} & = & 2\sum _{i_{1},i_{2},i_{3}=1}^{N}
\varepsilon _{i_{1}i_{2}i_{3}}
S_{0}^{(i_{1})}S_{1}^{(i_{2})}S_{2}^{(i_{3})}\\
\nonumber
&=&i^3 \sum _{i_{3}>i_{2}>i_{1}=1}^{N}
\left(z_{i_{1}i_{2}}z_{i_{2}i_{3}}z_{i_{3}i_{1}}
\partial _{z_{i_{3}}}\partial _{z_{i_{2}}}\partial _{z_{i_{1}}}+
sz_{i_{1}i_{2}}(z_{i_{2}i_{3}}-z_{i_{3}i_{1}})
\partial _{z_{i_{2}}}\partial _{z_{i_{1}}}
\right.\\ \nonumber &  & +  \left.
sz_{i_{2}i_{3}}(z_{i_{3}i_{1}}-z_{i_{1}i_{2}})
\partial _{z_{i_{3}}}\partial _{z_{i_{2}}} 
+sz_{i_{3}i_{1}}(z_{i_{3}i_{1}}-z_{i_{1}i_{2}})
\partial _{z_{i_{3}}}\partial _{z_{i_{2}}}
\right.\\
&& \left.  -  2s^{2}z_{i_{1}i_{2}}\partial _{z_{i_{3}}}
-2s^{2}z_{i_{2}i_{3}}\partial _{z_{i_{1}}}
-2s^{2}z_{i_{3}i_{1}}\partial _{z_{i_{2}}}\right)\,,
\lab{eq:q2q3}
\end{eqnarray}
where $z_{ij}=z_{i}-z_{j}$. Similar relations hold for 
the anti-holomorphic sector.

In that way one can also construct operators for the higher conformal charges.
They have a particularly simple form for the $\SL(2,\mathbb{C})$ spins $s=0$
\begin{equation}
\oq{k}=i^k\sum_{1\le j_1 < j_2 < \ldots  < j_k\le N}
z_{j_1j_2}\ldots z_{j_{k-1},j_k}z_{j_k,j_1}\partial_{z_{j_1}}\ldots 
\partial_{z_{j_{k-1}}}\partial_{z_{j_k}}
\lab{eq:qks0}
\end{equation}
as well as for $s=1$
\begin{equation}
\oq{k}=i^k\sum_{1\le j_1 < j_2 < \ldots  < j_k\le N}
\partial_{z_{j_1}}\ldots \partial_{z_{j_{k-1}}}\partial_{z_{j_k}}
z_{j_1j_2}\ldots z_{j_{k-1},j_k}z_{j_k,j_1}\,.
\lab{eq:qks1}
\end{equation}
The eigenvalues of the lowest conformal charge $\hat{q}_2$ can
be parameterized by a conformal weight $h$ (\ref{eq:hpar})
and complex spin $s$ (\ref{eq:spar}) as follows
\begin{equation}
q_2=-h(h-1)+Ns(s-1)
\lab{eq:q2}
\end{equation}

\subsection{Other symmetries}

The states (\ref{eq:Psip}) have additional symmetries \ci{Derkachov:2001yn}:
\begin{eqnarray}
\nonumber
 \mathbb{P}\Psi _{q,\wbar q}(\vec{z}_{1},\vec{z}_{2},\ldots,\vec{z}_{N}) & 
\eqdef & \Psi _{q,\wbar q}(\vec{z}_{2},\vec{z}_{3},\ldots,\vec{z}_{1})\\
\nonumber
&& = e^{i\theta_N (q,\wbar q)}\Psi _{q,\wbar q}(\vec{z}_{1},\vec{z}_{2},
\dots,\vec{z}_{N})\,,\\
\nonumber
 \mathbb{M}\Psi ^{\pm }(\vec{z}_{1},\vec{z}_{2},\ldots,\vec{z}_{N}) & 
\eqdef & \Psi ^{\pm }(\vec{z}_{N},\vec{z}_{N-1},\ldots,\vec{z}_{1}) \\ 
&& =  \pm \Psi ^{\pm }(\vec{z}_{1},\vec{z}_{2},\ldots,\vec{z}_{N})
\lab{eq:PMsym}
\end{eqnarray}
so called cyclic and mirror permutation
where the conformal charges are denoted by $q\equiv(q_2,q_3,\ldots,q_n)$ 
and $\wbar q\equiv(\wbar q_2,\wbar q_3,\ldots,\wbar q_n)$.
The generators  $\mathbb{P}$ and  $\mathbb{M}$, respectively, commute 
with the Hamiltonian $\cal H$ but they 
do not commute with each other. They satisfy  relations
\begin{equation}
 \mathbb{P}^N = \mathbb{M}^2=1, \quad
 \mathbb{P}^{\dagger} = \mathbb{P}^{-1}= \mathbb{P}^{N-1},  \quad
 \mathbb{M}^{\dagger} = \mathbb{M},  \quad
 \mathbb{P M} = \mathbb{M \, P}^{-1}= \mathbb{M \, P}^{N-1}\,.
\end{equation}

The phase $\theta_N (q)$ which is connected with eigenvalues of $\mathbb{P}$ 
is called quasimomentum. 
It takes the following values 
\begin{equation}
\theta_N(q,\wbar{q})=2 \pi \frac{k}{N}, \qquad \mbox{for } k=0,1,\ldots,N-1\,.
\lab{eq:quask}
\end{equation}
The eigenstates of the conformal
charges $\oq{k}$ diagonalize ${\cal H}$ and $\mathbb{P}$. 

The transfer matrices (\ref{eq:tnu}) are invariant under 
the cyclic permutations 
$\mathbb{P}^{\dagger}\, \ot{N}(u)\, \mathbb{P} = \ot{N}(u)$ whereas
they transform under the mirror transformation as
\begin{equation}
\mathbb{M}\, \ot{N}(u)\, \mathbb{M} = (-1)^N \ot{N}(-u).
\lab{eq:MtM}
\end{equation}
Substituting (\ref{eq:tnu}) into (\ref{eq:MtM}) one 
derives a transformation law of the conformal charges $q_k$
\begin{equation}
\mathbb{P}^{\dagger}\, \oq{k}\, \mathbb{P} = \oq{k}
\quad
\mathbb{M}\, \oq{k}\, \mathbb{M} = (-1)^k \oq{k}
\lab{eq:PMq}
\end{equation}
and similarly for the anti-holomorphic charges.
Since the Hamiltonian (\ref{eq:Ham}) is invariant under 
the mirror permutation, it has to satisfy 
\begin{equation}
{\cal H} (\oq{k},\oqb{k})=
\mathbb{M} {\cal H} (\oq{k},\oqb{k}) \mathbb{M}=
{\cal H} (\mathbb{M}\, \oq{k} \,\mathbb{M}, \mathbb{M}\, \oqb{k}\, \mathbb{M})=
{\cal H} ((-1)^k \oq{k},(-1)^k \oqb{k})\,.
\end{equation}
This implies that the eigenstates of the Hamiltonian (\ref{eq:Ham})
corresponding to two
different sets of the quantum number $\{q_k,\wbar{q}_k \}$ and
 $\{(-1)^k q_k,(-1)^k \wbar{q}_k \}$ have the same energy
\begin{equation}
E_N (q_k,\wbar{q}_k) = E_N ((-1)^k q_k,(-1)^k \wbar{q}_k)\,.
\end{equation}
Similarly one can derive relation for quasimomentum
\begin{equation}
\theta_N (q_k,\wbar{q}_k) = -\theta_N ((-1)^k q_k,(-1)^k \wbar{q}_k)\,.
\end{equation}

The cyclic and mirror permutation symmetries come
from the Bose symmetry and they appear
after performing the multi-colour limit \ci{'tHooft:1973jz}.
Physical states should possess both symmetries.

\section{Baxter $Q$-operator}
\subsection{Definition of the Baxter $Q$-operator}
The Schr\"odinger equation (\ref{eq:Schr}) may be solved applying the powerful method 
of the Baxter $Q$-operator \ci{Baxter}.
It depends on two complex spectral parameters $u$, $\wbar u$ and  
in the following will be denoted as
$\mathbb{Q}(u,\wbar{u})$.  
This operator has to satisfy the following relations
\begin{itemize}
\item Commutativity:
\begin{equation}
\left[ \mathbb{Q} (u,\wbar{u}) ,
 \mathbb{Q}(v,\wbar{v}) \right]=0\,,
\lab{eq:comQQ}
\end{equation}
\item $Q-t$ relation:
\begin{equation}
\left[\ot{N}(u), \mathbb{Q}(u,\wbar{u}) \right]=
\left[\otb{N}(\wbar{u}), \mathbb{Q}(u,\wbar{u}) \right]=0\,,
\lab{eq:comQt}
\end{equation}
\item Baxter equation:
\begin{equation}
\ot{N}(u) \mathbb{Q}(u,\wbar{u})  =
(u + i s)^N \mathbb{Q}(u+i,\wbar{u})  +
(u - i s)^N \mathbb{Q}(u-i,\wbar{u})  \,,
\lab{eq:Baxeq}
\end{equation}
\begin{equation}
\otb{N}(\wbar{u}) \mathbb{Q}(u,\wbar{u})  =
(\wbar{u} + i \wbar{s})^N \mathbb{Q}(u,\wbar{u}+i)  +
(\wbar{u} - i \wbar{s})^N \mathbb{Q}(u,\wbar{u}-i)  \,,
\lab{eq:Baxbeq}
\end{equation}
\end{itemize}
where $\ot{N}(u)$ and $\otb{N}(\wbar{u})$ are the auxiliary transfer matrices 
(\ref{eq:tnu}). 
According to (\ref{eq:comQt}) the Baxter $\mathbb{Q}(u,\wbar{u})$-operator
and the auxiliary transfer matrices as well as the Hamiltonian 
(\ref{eq:sepH}) share the common set of the eigenfunctions
\begin{equation}
 \mathbb{Q}(u,\wbar{u}) 
\Psi _{q,\wbar q}(\vec{z}_{1},\vec{z}_{2},\ldots,\vec{z}_{N}) 
=  Q_{q,\wbar q}(u,\wbar{u}) 
\Psi _{q,\wbar q}(\vec{z}_{1},\vec{z}_{2},\dots,\vec{z}_{N})\,.
\lab{eq:eeqQ}
\end{equation} 
The eigenvalues of the $Q$-operator satisfy the same Baxter equation
(\ref{eq:Baxeq}) and (\ref{eq:Baxbeq})
with the auxiliary transfer matrices replaced by their corresponding 
eigenvalues.

In the paper \ci{Derkachov:2001yn} the $Q$-operator was constructed 
as an $N-$fold integral operator
\begin{multline}
\mathbb{Q}(u,\wbar{u}) \Psi(\vec{z}_{1},\vec{z}_{2},\ldots,\vec{z}_{N}) =
\int d^2 w_1 \int d^2 w_2 
\ldots \int d^2 w_N \\
Q_{u,\wbar{u}} 
(\vec{z}_{1},\vec{z}_{2},\ldots,\vec{z}_{N} |
\vec{w}_{1},\vec{w}_{2},\ldots,\vec{w}_{N})
\Psi(\vec{w}_{1},\vec{w}_{2},\ldots,\vec{w}_{N}) \,,
\lab{eq:convQ}
\end{multline}
where the integrations are performed over two-dimensional
$\vec{w}_i-$planes.
The integral kernel in (\ref{eq:convQ}) takes two different
forms:
\begin{multline}
Q^{(+)}_{u,\wbar{u}}(\vec{z}|\vec{w})=\\
=a(2-2s,s+iu,\wbar{s}-i\wbar{u})^N {\pi}^N
\prod_{k=1}^{N}
\frac{\left[z_k-z_{k+1} \right]^{1-2s}}
{\left[w_k-z_k \right]^{1-s-iu}
\left[w_k-z_{k+1} \right]^{1-s+iu}}
\lab{eq:Qpker}
\end{multline}
and
\begin{equation}
Q^{(-)}_{u,\wbar{u}}(\vec{z}|\vec{w})=
\prod_{k=1}^{N}
\frac{\left[w_k-w_{k+1} \right]^{2s-2}}
{\left[z_k-w_k \right]^{s+iu}
\left[z_k-w_{k+1} \right]^{s-iu}}\,,
\lab{eq:Qmker}
\end{equation}
which appear to be equivalent \ci{Derkachov:2001yn}.
In Eqs. (\ref{eq:Qpker}) and (\ref{eq:Qmker}) 
function $a(\ldots)$ factorizes as 
\begin{equation}
a(\alpha,\beta,\ldots)=a(\alpha)a(\beta)\ldots 
\quad
\mbox{and} 
\quad
a(\alpha)=\frac{\Gamma(1-\wbar{\alpha})}{\Gamma(\alpha)}
\lab{eq:aalph}
\end{equation}
and
$\wbar{\alpha}$ is an anti-holomorphic partner of $\alpha$ satisfying 
$\alpha- \wbar{\alpha} \in \mathbb{Z}$.
Moreover, the two-dimensional propagators are defined as 
\begin{equation}
\left[z_k-w_k \right]^{-\alpha}=
\left(z_k-w_k \right)^{-\alpha}
\left(\wbar{z}_k-\wbar{w}_k \right)^{-\wbar{\alpha}}\,.
\lab{eq:2dprop}
\end{equation}
In order for the Baxter $Q-$operators to be well defined, 
(\ref{eq:Qpker}) and (\ref{eq:Qmker}), should be  single-valued functions.
In this way we can find that the spectral parameters $u$ and $\wbar{u}$
have to satisfy the condition
\begin{equation}
i(u-\wbar{u})=n
\lab{eq:uubarn}
\end{equation}
with $n$ being an integer.

The Baxter $Q$-operator has a well defined pole structure. For  
$\mathbb{Q}^{(+)}(u,\wbar{u})$ we have an infinite set of poles
of the order not higher than $N$ situated at
\begin{equation}
\left\{\ u_m^+=i(s-m), \wbar{u}_{\wbar{m}}^+=i(\wbar{s}-\wbar{m})\right\};
\quad
\left\{\ u_m^-=-i(s-m), \wbar{u}_{\wbar{m}}^-=-i(\wbar{s}-\wbar{m})\right\}
\lab{eq:upoles}
\end{equation}
with  $m,\wbar m=1,2,\ldots \,$.
The behaviour of 
$Q_{q_,\wbar{q}}(u,\wbar{u})\equiv Q^{(+)}_{q_,\wbar{q}}(u,\wbar{u})$ 
\ie an eigenvalue of $\mathbb{Q}^{(+)}(u,\wbar{u})$,  
in the vicinity of the pole at $m=\wbar
m=1$ can be parameterized as
\begin{equation}
Q_{q,\wbar q}(u_{1}^{\pm}+\epsilon,{\wbar u}_{1}^{\pm}+\epsilon)=R^\pm(q,\wbar
q)\left[\frac1{\epsilon^N} +\frac{i\,E^\pm(q,\wbar q)}{\epsilon^{N-1}}+ \ldots 
\,\right]\!.
\lab{eq:Q-R,E}
\end{equation}
\\[-3mm]
The functions $R^\pm(q,\wbar q)$ fix an overall normalization of the Baxter
operator, while the residue functions $E^\pm(q,\wbar q)$ define the energy of the
system (see Eqs.~(\ref{eq:RR}) and (\ref{eq:energy}) below).
It has also specified asymptotic behaviour. For $|\IIm \lambda| < 1/2$
and $\RRe \lambda \rightarrow \infty$
\begin{equation}
Q_{q,\wbar q}(\lambda-i n/2,\lambda+i n/2) \sim
\e^{i \Theta_h(q,\wbar q)} \lambda^{h +\wbar h-N(s-\wbar s)}+
\e^{-i \Theta_h(q,\wbar q)} \lambda^{1-h +1-\wbar h-N(s-\wbar s)} \,,
\lab{eq:Qanalb}
\end{equation}
where $\Theta_h$ is a phase that should not be confused 
with quasimomentum $\theta_N(q,\wbar q)$.

\subsection{Observables}

The Hamiltonian (\ref{eq:sepH}) may be written in terms of 
the Baxter $Q$-operator \ci{Derkachov:2001yn}:
\begin{multline}
{\cal H}_N=\epsilon_N + \left. i \frac{d}{du} 
\ln \mathbb{Q}^{(+)}(u+is,\wbar{u}+i \wbar{s})\right|_{u=0}\\
-\left( \left. i \frac{d}{du} 
\ln \mathbb{Q}^{(+)}(u-is,\wbar{u}-i \wbar{s})\right|_{u=0}\right)^{\dagger}\,,
\lab{eq:HNQQ}
\end{multline}
where the additive normalization constant is given as
\begin{equation}
\epsilon_N = 2 N \, \RRe [\psi(2 s)+ \psi(2-2s)-2 \psi(1)]\,.
\lab{eq:epsnor}
\end{equation}

Applying to (\ref{eq:HNQQ}) the eigenstate $\Psi_q$ we obtain the
energy 
\begin{equation}
E_N(q,\bar q)=\varepsilon_N + i\frac{d}{du} \ln\left[ Q_{q,\bar
q}^{}(u+is,u+i\bar s)\, \left(Q_{q,\bar q}^{}(u-is,u-i\bar
s)\right)^*\right]\bigg|_{u=0},
\lab{eq:enQ}
\end{equation}
or equivalently
\begin{multline}
E_N(q,\bar q) = - \Im\frac{d}{du}\ln \bigg[u^{2N}Q_{q,\bar
q}(u+i(1-s),u+i(1-\bar s))\,\\
\times Q_{-q,-\bar q}(u+i(1-s),u+i(1-\bar s))\bigg]\bigg|_{u=0} \,,
\lab{eq:enQ2}
\end{multline}
where
$Q_{q_,\wbar{q}}(u,\wbar{u})\equiv Q^{(+)}_{q_,\wbar{q}}(u,\wbar{u})$ 
is eigenvalue
of the $ \mathbb{Q}^{(+)}(u,\wbar{u})$ operator,
while 
\begin{equation}
\pm q=(q_2,\pm q_3,\ldots,(\pm)^N q_N) 
\lab{eq:pmq}
\end{equation}
are 
the conformal charges.

It is also possible to rewrite the quasimomentum operator 
in terms of  $ \mathbb{Q}_+(u,\wbar{u})$:
\begin{equation}
  \Hat{{\theta}}_N=-i \ln \mathbb{P} 
= i \ln \frac{ \mathbb{Q}_{+}(is,i\wbar{s})}{
\mathbb{Q}_{+}(-is,-i\wbar{s})}\,.
\lab{eq:quasQ}
\end{equation}
Moreover, using the mirror permutation (\ref{eq:PMq}) one finds 
the following parity relations for the residue functions $R^+(q,\wbar q)$
defined in (\ref{eq:Q-R,E}):
\begin{equation}
R^+(q,\wbar q)/R^+(-q,-\wbar q)=\e^{2i\theta_N(q,\wbar q)}
\lab{eq:RR}
\end{equation}
and for the eigenvalues of the Baxter operator:
\begin{equation}
Q_{q,\wbar q}(-u,-\wbar u) = \e^{i\theta_N(q,\wbar q)} 
Q_{-q,-\wbar q}(u,\wbar u)\,,
\lab{eq:Q-symmetry}
\end{equation}
where $-q\equiv(q_2,-q_3,\ldots ,(-1)^n q_n)$ and similarly for $\wbar q$. 
Examining the behaviour of (\ref{eq:Q-symmetry}) 
around the pole
at $u=u_{1}^{\pm}$ and $\wbar u={\wbar u}_{1}^{\pm}$ and making use of
Eq.~(\ref{eq:Q-R,E}) one gets
\begin{equation}
R^\pm(q,\wbar q)=(-1)^N\e^{i\theta_N(q,\wbar q)}R^\mp(-q,-\wbar q)\,,\qquad
E^\pm(q,\wbar q)=-E^\mp(-q,-\wbar q)\,.
\lab{eq:sym}
\end{equation}

To obtain the expression for the energy $E_N(q,\wbar q)$, we apply 
(\ref{eq:enQ})
and
replace the function $Q_{q,\wbar q}(u\pm i(1-s),u\pm i(1-\wbar
s))$ by its pole expansion (\ref{eq:Q-R,E}). 
Then, applying the second relation in
(\ref{eq:sym}), one finds
\begin{equation}
E_N(q,\wbar q)= E^+(-q,-\wbar q)+ (E^+(q,\wbar q))^*=\Re\left[E^+(-q,-\wbar q)+
E^+(q,\wbar q)\right]\,,
\lab{eq:energy}
\end{equation}
where the last relation follows from hermiticity of the Hamiltonian 
(\ref{eq:Ham}). We
conclude from Eqs.~(\ref{eq:energy}) and (\ref{eq:Q-R,E}), 
that in order to find the energy
$E_N(q,\wbar q)$, one has to calculate the residue of 
$Q_{q,\wbar q}(u,\wbar u)$ at
the $(N-1)$-th order pole at $u=i(s-1)$ and $\wbar u=i(\wbar s-1)$.

\subsection{Construction of the eigenfunction}

The Hamiltonian eigenstate $\Psi_{\vec{p},\{q,\wbar{q}\}}(\vec{z})$
is a common eigenstate of the total set of the integrals of motion,
$\vec{p}$ and $\{q,\wbar{q}\}$ as well as the Baxter $Q$-operator.
Thanks to the method of the Separation of Variables (SoV) developed by
Sklyanin \ci{Sklyanin:1991ss,Derkachov:2001yn} we can write the eigenstate  
using separated coordinates
$\vec{\mybf x}=(\vec{x}_1,\ldots,\vec{x}_{N-1})$ as
\begin{multline}
\Psi_{\vec{p},\{q,\wbar{q}\}}(\vec{z})
=\int d^{N-1} \vec{\mybf x} \,
\mu(\vec{x}_1,\ldots,\vec{x}_{N-1})
\, U_{\vec{p},\vec{x}_1,\ldots,\vec{x}_{N-1}}(\vec{z}_1,\ldots,\vec{z}_{N})\\
\times \left(\Phi_{q,\wbar{q}}(\vec{x}_1,\ldots,\vec{x}_{N-1})\right)^{\ast}\,,
\lab{eq:PsiU}
\end{multline}
where $U_{\vec{p},\vec{\mybf x}}$ is the kernel of the unitary operator while
\begin{multline}
\left(\Phi_{q,\wbar{q}}(\vec{x}_1,\ldots,\vec{x}_{N-1})\right)^{\ast}=\\
=e^{i \theta_N(q,\wbar{q})/2}
\prod_{k=1}^{N-1}
\left(
\frac{\Gamma(s+ix_k)\Gamma(\wbar{s}-i\wbar{x}_k)}{
\Gamma(1-s+ix_k)\Gamma(1-\wbar{s}-i\wbar{x}_k)} 
\right)^N
Q_{q,\wbar{q}}(x_k,\wbar{x}_k)\,.
\lab{eq:PhiQ}
\end{multline}
The functions
$Q_{q,\wbar{q}}(x_k,\wbar{x}_k)$ are eigenstates of the Baxter $Q$-operator.
In contrast to the $\vec{z}_i=(z_i,\wbar{z}_i)$ - coordinates, 
the allowed values
of separated coordinates are
\begin{equation}
x_k=\nu_k-\frac{i n_k}{2}\,, \qqquad
\wbar{x}_k=\nu_k+\frac{i n_k}{2}
\lab{eq:xcoords}
\end{equation} 
with $n_k$ integer and $\nu_k$ real. Integration over the space of separated
variables implies summation over integer $n_k$ and integration over 
continuous $\nu_k$
\begin{equation}
\int d^{N-1} \vec{\mybf x}=
\prod_{k=1}^{N-1} 
\left(\sum_{n_k=-\infty}^{\infty} \int_{-\infty}^{\infty} d \nu_k \right),
\quad
\mu(\vec{\mybf x})=\frac{2 \pi^{-N^2}}{(N-1)!}
\prod_{\shortstack{$\scriptstyle j,k=1$\\$\scriptstyle j>k$}}^{N-1}
\left| \vec{x}_k - \vec{x}_j \right|^2 \,,
\lab{eq:xmes}
\end{equation}
where $\left| \vec{x}_k - \vec{x}_j \right|^2=(\nu_k-\nu_j)^2+
(n_h-n_j)^2/4$.

The integral kernel  $U_{\vec{p},\vec{\mybf x}}$ can be written as
\begin{equation}
U_{\vec{p},\vec{\mybf x}}(\vec{z}_1,\ldots,\vec{z}_N)
=c_N(\vec{\mybf x})(\vec{p}^{\, \,2})^{(N-1)/2}
\int d^2 w_N e^{2 i \vec{p} \cdot \vec{w}_N}
U_{\vec{\mybf x}}(\vec{z}_1,\ldots,\vec{z}_N;\vec{w}_N)\,,
\lab{eq:Upker}
\end{equation}
where $2 \vec{p} \cdot \vec{w}_N=p \, w_N+ \wbar{p} \, \wbar{w}_N$,
\begin{equation}
U_{\vec{\mybf x}}(\vec{z}_1,\ldots,\vec{z}_N;\vec{w}_N)=
\left[
\Lambda_{N-1,(\vec{x}_1)}^{(s,\wbar{s})}
\Lambda_{N-2,(\vec{x}_2)}^{(1-s,1-\wbar{s})}
\ldots
\Lambda_{1,(\vec{x}_{N-1})}^{(s,\wbar{s})}
\right](\vec{z}_1,\ldots,\vec{z}_N|\vec{w}_N)
\lab{eq:Uweker}
\end{equation}
for even $N$, and
\begin{equation}
U_{\vec{\mybf x}}(\vec{z}_1,\ldots,\vec{z}_N;\vec{w}_N)=
\left[
\Lambda_{N-1,(\vec{x}_1)}^{(s,\wbar{s})}
\Lambda_{N-2,(\vec{x}_2)}^{(1-s,1-\wbar{s})}
\ldots
\Lambda_{1,(\vec{x}_{N-1})}^{(1-s,1-\wbar{s})}
\right](\vec{z}_1,\ldots,\vec{z}_N|\vec{w}_N)
\lab{eq:Uwoker}
\end{equation}
for odd $N$. Here the convolution involves the product of $(N-1)$
functions $\Lambda_{N-k,(\vec{x}_k)}$ with alternating spins $(s,\wbar{s})$
and $(1-s,1-\wbar{s})$. They are defined as
\begin{multline}
\Lambda_{N-n,(\vec{x})}^{(s,\wbar s)}
(\vec{z}_n,\ldots,\vec{z}_N|\vec{y}_{n+1},\ldots,\vec{y}_N)=
\left[z_1-y_2\right]^{-x+iu} \\
\times \left(
\prod_{k=n+1}^{N-1}
\left[z_k-y_k\right]^{-x-iu}
\left[z_k-y_{k+1}\right]^{-x+iu}
\right)
\left[z_N-y_N\right]^{-x-iu}\,,
\lab{eq:Lambker}
\end{multline}
where the convolution  $[\Lambda_{N-k,(\vec{x}_{k})}
\Lambda_{N-k+1,(\vec{x}_{k-1})}]$
contains $(N-k)$ two-di\-men\-sional integrals.
The coefficient $c_N(\vec{\mybf x})$  is given for $N \ge 3$
\begin{multline}
 c_N(\vec{\mybf x})=
\prod_{k=1}^{[(N-1)/2]}
\left( a(s+ix_{2k},\wbar{s}-i\wbar{x}_{2k})\right)^{N-k}\\
\times \prod_{k=1}^{[N/2-1]}
\left( a(s+ix_{2k+1},\wbar{s}-i\wbar{x}_{2k+1})\right)^{k}
\lab{eq:cnx}
\end{multline}
while the products go over integer numbers lower than upper limit.
For $N=2$ we have $c_2(\vec{x}_1)=1$.

\section{Quantization conditions in the $Q$-Baxter method}

In Ref. \ci{Derkachov:2002wz}
the authors describe a construction of 
the solution to the Baxter equations, 
(\ref{eq:Baxeq}) and (\ref{eq:Baxbeq}),
which satisfies additionally the conditions (\ref{eq:upoles}) 
and (\ref{eq:Qanalb}).
This can be done by means of
the following integral representation for $Q_{q,\wbar q}(u,\wbar u)$
\begin{equation}
Q_{q,\wbar q}(u,\wbar u)= \int\frac{d^2 z}{z\wbar z}\, 
z^{-i u} {\wbar z}^{-i\wbar u}\, Q(z,\wbar z)\,,
\lab{eq:Q-R}
\end{equation}
where we integrate over the two-dimensional 
$\vec z-$plane with $\wbar z=z^*$
and $Q(z,\wbar z)$ depends on $\{q,\wbar q\}$.
The advantages of this ansatz are:
\begin{itemize}
\item the functional Baxter equation on $Q_{q,\wbar q}(u,\wbar u)$ is
transformed into the $N-$th order differential equation for the function  
$Q(z,\wbar z)$
\begin{multline}
\left[z^s\lr{z\partial_z}^{N}z^{1-s}+z^{-s}\lr{z\partial_z}^{N}z^{s-1}
\right. \\ \left.
-2\lr{z\partial_z}^{N}-\sum_{k=2}^N i^{k}q_k\lr{z\partial_z}^{N-k}
\right]Q(z,\wbar z)=0\,.
\lab{eq:Eq-1}
\end{multline}
A similar equation holds in the anti-holomorphic sector 
with $s$ and $q_k$ replaced by $\wbar s=1-s^*$ and
$\wbar q_k=q_k^*$, respectively. 
\item 
the condition (\ref{eq:uubarn})
is automatically satisfied since 
the $z-$integral in the r.h.s.\ of (\ref{eq:Q-R}) 
is well-defined 
only for 
$i(u-\wbar u)=n$.
\item the remaining two
conditions for the analytical properties and asymptotic behaviour of 
$Q_{q,\wbar q}(u,\wbar u)$, Eqs.~(\ref{eq:upoles}) and (\ref{eq:Qanalb}), 
become equivalent to the
requirement for $Q(z,\wbar z=z^*)$ to be a single-valued function 
on the complex
$z-$plane.
\end{itemize}

The differential equation (\ref{eq:Eq-1})
is of Fuchsian type. It possesses three regular singular points located 
at $z=0$, $z=1$ and $z=\infty$.
Moreover, it has
$N$ linearly independent solutions,
$Q_a(z)$. 
The anti-holomorphic equation has also
$N$ independent solutions,
$\wbar Q_b(\wbar z)$.

Now,
we construct the general expression for the
function $Q(z,\wbar z)$ as
\begin{equation}
Q(z,\wbar z) = \sum_{a,b=1}^N Q_a(z)\, C_{ab}\, \wbar Q_b(\wbar z)\,,
\lab{eq:general-sol}
\end{equation}
where $C_{ab}$ is an arbitrary mixing matrix. The functions $Q_a(z)$ and
$\wbar Q_b(\wbar z)$ 
have a nontrivial monodromy\footnote{The monodromy matrix around 
$z=0$ is defined as $Q_n^{(0)}(z\e^{2\pi i})=M_{nk}Q_k^{(0)}(z)$
and 
similarly for the other singular points.} 
around three singular points, 
$z,\,\wbar z=0$, $1$
and $\infty$. 
In order to be well-defined on the whole plane, functions $Q(z,\wbar z=z^*)$ 
should be single-valued and
their
monodromy should cancel in the r.h.s.\ of (\ref{eq:general-sol}). 
This condition allows us to determine the values of the mixing coefficients,
$C_{ab}$, and also to calculate the quantized values of the 
conformal charges $q_k$.

The differential equation (\ref{eq:Eq-1}) is also 
symmetric under the  transformation
$z\to 1/z$ and $q_k\to (-1)^k q_k$. 
This property is related to 
Eq.~(\ref{eq:Q-symmetry}) and leads to
\\[1mm]
\begin{equation}
Q_{q,\wbar q}(z,\wbar z)=\e^{i\theta_N(q,\wbar q)}
Q_{-q,-\wbar q}(1/z,1/\wbar z)\,,
\lab{eq:Qz-symmetry}
\end{equation}
\\[0mm]
where $\pm q=(q_2,\pm q_3,\ldots,(\pm)^N q_N)$ denotes 
the integrals of motion corresponding to the function $Q(z,\wbar{z})$.
The above formula allows us to define the solution 
$Q(z,\wbar{z})$ around 
$z=\infty$ from the solution at $z=0$. Thus, applying (\ref{eq:Qz-symmetry})
we are able to find $Q(z,\wbar{z})$ and analytically continue it to the whole 
$z-$plane.

\subsection{Solution around $z=0$}

We find  a solution $Q(z)\sim z^a$ by the series method.
The indicial equation for the solution of Eq. (\ref{eq:Eq-1})
around $z=0$ reads as follows
\begin{equation}
(a-1+s)^N=0
\lab{eq:indicial-0}
\end{equation}
and the solution, $a=1-s$ is $N-$fold degenerate. This leads to
terms $\sim \Log^k(z)$ with $k\le N-1$.
We define the
fundamental set of linearly independent solutions to (\ref{eq:Eq-1}) 
around $z=0$ as
\begin{eqnarray}
Q_1^{(0)}(z)&=& z^{1-s} u_1(z)\,,
\nonumber\\ \nonumber
Q_m^{(0)}(z)&=&z^{1-s}\left[u_1(z) \Log^{m-1}(z)
\right. \\ 
& & +\left.\sum_{k=1}^{m-1}c_{m-1}^k u_{k+1}(z) \Log^{m-k-1}(z)\right],
\lab{eq:Q-0-h}
\end{eqnarray}
with $2\le m\le N$ and where for the later
convenience 
\begin{equation}
c_{m-1}^k=\frac{(m-1)!}{(k!(m-k-1)!)}.
\end{equation}
The functions $u_m(z)$ are defined
inside the region $|z|<1$
and have a form
\begin{equation}
u_m(z) = 1+\sum_{n=1}^\infty z^n\,u^{(m)}_{n}(q)\,.
\lab{eq:power-series-0}
\end{equation}
Inserting (\ref{eq:Q-0-h}) and
(\ref{eq:power-series-0}) into (\ref{eq:Eq-1}), 
one derives recurrence relations for $u^{(m)}_n(q)$.
However, in order to save space, we do not show here their explicit form.

In the anti-holomorphic sector the 
fundamental set of solutions 
can be obtained from (\ref{eq:Q-0-h}) 
by substituting $s$
and $q_k$ by $\wbar s=1-s^*$ and $\wbar q_k=q_k^*$, respectively. 
Sewing the two sectors we obtain
 the general
solution for $Q(z,\wbar z)$ around $z=0$ as
\begin{equation}
Q(z,\wbar z) \stackrel{|z|\to 0}{=} \sum_{m,\wbar m=1}^N Q^{(0)}_m(z)\,
C^{(0)}_{m\wbar m}\,\wbar{Q}^{(0)}_{\wbar m}(\wbar z)\,.
\lab{eq:Q-0}
\end{equation}
The above solution (\ref{eq:Q-0}) should be single-valued on the $z-$plane.
Thus, imposing single-valuedness condition on (\ref{eq:Q-0}) 
we find a structure of the mixing matrix $C^{(0)}_{m\wbar m}$
which for $n+m\le N+1$
\begin{equation}
C^{(0)}_{nm}=\frac{\sigma}{(n-1)!(m-1)!}
\sum _{k=0}^{N-n-m+1}{\frac {(-2)^{k}}{k!}\,\alpha_{k+n+m-1}}
\lab{eq:C0}
\end{equation}
with $\sigma, \alpha_1,\ldots ,\alpha_{N-1}$ being arbitrary 
complex parameters and $\alpha_N=1$. Below the main anti-diagonal,
that is for $n+m> N+1$, $C^{(0)}_{nm}$ vanish.

The mixing matrix $C^{(0)}_{m\wbar m}$ depends on $N$ arbitrary complex parameters
$\sigma$ and $\alpha_k$. 
However, two parity relations, Eqs.~(\ref{eq:RR}) and (\ref{eq:Qz-symmetry}), 
fix $\sigma=\exp(i\theta_N(q,\wbar q))$, 
with $\theta_N(q,\wbar q)$ being
the quasimomentum,
and lead to
the quantization of the quasimomentum. 
Later, we will use 
(\ref{eq:Qz-symmetry}) to
calculate the eigenvalues of $\theta_N(q,\wbar q)$ 
(see Eq.~(\ref{eq:parity-qc})).

The leading
asymptotic behaviour of $Q(z,\wbar{z})$ 
for $z\to 0$ can be obtained 
by substituting (\ref{eq:C0}) and (\ref{eq:Q-0-h}) into (\ref{eq:Q-0}). 
It has a form
\begin{multline}
Q_{q,\wbar q}(z,\wbar z)=z^{1-s}\wbar z^{1-\wbar s}\e^{i\theta_N(q,\wbar
q)}\left[\frac{\Log^{N-1} (z\wbar z)}{(N-1)!} +
\frac{\Log^{N-2} (z\wbar z)}{(N-2)!}\, \alpha_{N-1}+ \right.\\
\left. \ldots  +
\frac{\Log (z\wbar z)}{1!}\, \alpha_{2}+\alpha_1\right]
\lr{1+{\cal O}(z,\wbar z)}\,.
\lab{eq:Q-small}
\end{multline}
Making use of the integral identity 
\begin{multline}
\int_{|z|<\rho}\frac{d^2 z}{z\bar z} z^{-iu}\bar z^{-i\bar u}
\ln^n(z\bar z) z^{m-s} \bar z^{\bar m-\bar s}=\\
=\pi\delta_{m-s-iu,\bar m-\bar s-i\bar u}\left[
\frac{(-1)^n\,n!}{(m-s-iu)^{n+1}}+{\cal O}((m-s-iu)^0)\right],
\lab{eq:intident}
\end{multline}
with $m$ and $\wbar m$ positive integer,
we can calculate the contribution of the small$-z$ region to the eigenvalue 
of the Baxter equation (\ref{eq:Q-R}). The
function $Q_{q,\wbar q}(u,\wbar u)$ has poles of the order $N$ in the points
$u=i(s-m)$ and $\wbar u=i(\wbar s-\wbar m)$ what agrees with (\ref{eq:upoles}). 
For $m=\wbar m=1$ one finds from
(\ref{eq:Q-small})
\begin{multline}
Q_{q,\wbar q}(u_{1}^{+}+\epsilon,{\wbar u}_{1}^{+}+\epsilon)=\\
=-\frac{\pi
\e^{i\theta_N(q,\wbar
q)}}{(i\epsilon)^N}
\left[1+i\epsilon\,\alpha_{N-1}+\ldots +(i\epsilon)^{N-2}
\,\alpha_2+(i\epsilon)^{N-1}\,\alpha_1+{\cal O}(\epsilon^N)
\right]\,,
\lab{eq:Q-pole-0}
\end{multline}
where $u_{1}^{+}$ and ${\wbar u}_{1}^{+}$ are defined in (\ref{eq:upoles}). 
One can see that
the integration in (\ref{eq:Q-R}) over the region of large $z$
with (\ref{eq:Qz-symmetry}) and (\ref{eq:Q-small}) 
gives the second set of poles for  $Q_{q,\wbar q}(u,\wbar u)$ 
located at $u=-i(s-m)$ and $\wbar u=-i(\wbar s-\wbar m)$.

Comparing (\ref{eq:Q-pole-0}) with (\ref{eq:Q-R,E}) one obtains
\begin{equation}
R^+(q,\wbar q)=-\frac{\pi}{i^N}\e^{i\theta_N(q,\wbar q)}\,,\qquad E^+(q,\wbar
q)=\alpha_{N-1}(q,\wbar q)\,.
\lab{eq:E=alpha}
\end{equation}
Now, we may derive expression for the energy 
\begin{equation}
E_N(q,\wbar q)=\Re\left[\alpha_{N-1}(-q,-\wbar q)+\alpha_{N-1}(q,\wbar q)\right]\,.
\lab{eq:E-fin}
\end{equation}
The arbitrary complex parameters $\alpha_n$, defined in (\ref{eq:C0}),
will be fixed by 
the quantization conditions below.

In this Section we have obtained following Ref. \ci{Derkachov:2002wz} the 
expression for the energy spectrum 
$E_N(q,\wbar q)$, as a function of the matrix
elements of the mixing matrix (\ref{eq:C0}) in the fundamental basis 
(\ref{eq:Q-0-h}). Moreover, we have defined the solution to the 
Baxter equation
$Q(u,\wbar{u})$ and reproduced the analytical properties
of the eigenvalues of the Baxter operator, 
Eq.~(\ref{eq:upoles}).

\subsection{Solution around $z=1$}

Looking for a solution of (\ref{eq:Eq-1}) 
around $z=1$ in a form $Q(z)\sim(z-1)^b$
we obtain the following indicial equation
\begin{equation}
(b+1+h-Ns)(b+2-h-Ns)\prod_{k=0}^{N-3}(b-k)=0\,,
\lab{eq:b-exponents}
\end{equation}
where $h$ is the total $\SL(2,\mathbb{C})$ spin  defined in (\ref{eq:q2}).
Although the solutions $b=k$ with $k=0,\ldots ,N-3$ differ from each other 
by an integer, for $h\neq (1+n_h)/2$, no logarithmic terms appear.
The $\Log(z)-$terms are only needed for $\IIm h = 0$ where the additional 
degeneration occurs. 

Thus, we define 
the fundamental set of solutions to Eq.~(\ref{eq:Eq-1}) around $z=1$. 
For $\IIm h\neq 0$ it has the form
\begin{eqnarray}
&&Q_1^{(1)}(z) = z^{1-s} (1-z)^{Ns-h-1}v_1(z)\,,
\nonumber
\\[2mm]
&&Q_2^{(1)}(z) = z^{1-s} (1-z)^{Ns+h-2}v_2(z)\,,
\nonumber
\\[2mm]
&&Q_m^{(1)}(z) = z^{1-s} (1-z)^{m-3} v_m(z)\,,
\lab{eq:set-1}
\end{eqnarray}
with $m=3,\ldots ,N$. The functions $v_{i}(z)$ $(i=1,2)$ and $v_m(z)$ given by the
power series
\begin{equation}
v_i(z)=1+\sum_{n=1}^\infty (1-z)^n \,v^{(i)}_{n}(q)\,,
\qquad
v_m(z)=1+\sum_{n=N-m+1}^\infty (1-z)^n\, v^{(m)}_{n}(q)\,,
\lab{eq:v-series}
\end{equation}
which converge inside the region $|1-z|<1$
and where 
the expansion coefficients
 $v^{(i)}_{n}$ and $v^{(m)}_{n}$ satisfy the $N-$term 
recurrence relations\footnote{
The factor $z^{1-s}$ was included in the r.h.s.\ of 
(\ref{eq:set-1}) and (\ref{eq:Q-deg})
to simplify the form of the recurrence relations.}
with respect to the index $n$.
For $h=(1+n_h)/2 \in 2 \mathbb{Z}+1$, one $\Log(z)-$terms appear
so for
 $n_h\ge 0$, 
\begin{multline}
Q_1^{(1)}(z)\bigg|_{h=(1+n_h)/2} = \\
=z^{1-s} (1-z)^{Ns-(n_h+3)/2} \left[(1-z)^{n_h}
\Log (1-z)\, v_2(z)
+ \widetilde v_1(z)\right]\,,
\lab{eq:Q-deg}
\end{multline}
where the function $v_2(z)$ is the same as before, $\widetilde
v_1(z)=\sum_{k=0}^\infty\tilde v_k z^k$ and the coefficients $\tilde v_k$ 
satisfy recurrence relations with the boundary condition 
$\tilde v_{n_h}=1$.
For $h \in \mathbb{Z}$ we have two additional terms:
$\Log(z)$ and $\Log^2(z)$.

Similar calculations have to be performed in the anti-holomorphic sector
with $s$ and $h$ replaced by
$\wbar s=1-s^*$ and $\wbar h=1-h^*$, respectively. 
A general solution for $Q(z,\wbar{z})$ for $\IIm(h) \ne 0$
with respect to the single-valuedness 
can be constructed as 
\begin{multline}
Q(z,\wbar z)\stackrel{|z|\to 1}{=}\beta_h Q_1^{(1)}(z)\wbar Q_1^{(1)}(\wbar
z)+\beta_{1-h} Q_2^{(1)}(z)\wbar Q_2^{(1)}(\wbar z) \\
+\sum_{m,\wbar m=3}^N
Q_m^{(1)}(z)\,\gamma_{m\wbar m}\,\wbar Q_{\wbar m}^{(1)}(\wbar z)\,.
\lab{eq:Q-1}
\end{multline}
Here
the parameters $\beta_h$ and $\gamma_{m \wbar m}$ build the $C^{(1)}$ matrix
where $Q(z,\wbar{z})=Q^{(1)}_m C^{(1)}_{m\wbar{m}} \wbar{Q}^{(1)}_{\wbar{m}}$.
The $\beta-$coefficients depend, in general, on the total spin $h$ (and
$\wbar h=1-h^*$). They are chosen in (\ref{eq:Q-1}) 
in such a way that the symmetry of
the eigenvalues of the Baxter operator under $h\to 1-h$ becomes manifest. 
Thus, the mixing matrix $C^{(1)}$ defined in (\ref{eq:Q-1}) 
depends on $2+(N-2)^2$ complex parameters $\beta_h$, $\beta_{1-h}$
and $\gamma_{m\wbar m}$ which are some functions 
of the integrals of
motion $(q,\wbar q)$, so, they can be fixed by the quantization conditions. 

For $h=(1+n_h)/2$ the first two terms in the
r.h.s.\ of (\ref{eq:Q-1}) look differently in virtue of (\ref{eq:Q-deg})
\begin{multline}
Q(z,\wbar z)\bigg|_{h=(1+n_h)/2}=\beta_1\! \left[Q_1^{(1)}(z)\wbar
Q_2^{(1)}(\wbar z)+Q_2^{(1)}(z)\wbar Q_1^{(1)}(\wbar z)\right]\\
+\beta_2\,
Q_2^{(1)}(z)\wbar Q_2^{(1)}(\wbar z) + \ldots \, ,
\lab{eq:Q-1l}
\end{multline}
where ellipses denote the remaining terms\footnote{Equation (\ref{eq:Q-1l})
describes solutions only for $h=(1+n_h)/2$ where $n_h \in 2 \mathbb{Z}$.}. 

Substituting 
(\ref{eq:Q-1}) and (\ref{eq:Q-1l}) into (\ref{eq:Q-R})
and performing integration over the region of $|1-z|\ll 1$, one can find the
asymptotic behaviour of $Q(u,\wbar u)$ at large $u$. 

Let us consider the duality relation
(\ref{eq:Qz-symmetry}). 
Using the function $Q(z,\wbar{z})$  
we evaluate (\ref{eq:Q-1}) in the limit $|z|\rightarrow1$. 
In this way, we obtain set of relations for
the functions
$\beta_i(q,\wbar q)$ and $\gamma_{m\wbar m}(q,\wbar q)$. 
The derivation is based
on the following property 
\begin{equation}
Q^{(1)}_a(1/z;-q)= \sum_{b=1}^N S_{ab} \,Q^{(1)}_b(z;q)\,,
\lab{eq:S-def}
\end{equation}
with $\Im(1/z)>0$
and where the dependence on the integrals of motion
was explicitly indicated.
Here taking limit $z\rightarrow 1$ in (\ref{eq:set-1}) and 
(\ref{eq:v-series}) and substituting them to (\ref{eq:S-def}) 
we are able to
evaluate the $S-$matrix
\begin{equation}
S_{11}=\e^{-i\pi(Ns-h-1)}\,,\;S_{22}=\e^{-i\pi(Ns+h-2)}
\,,\;
S_{k,k+m}=(-1)^{k-3}\frac{(k-2s-1)_m}{m!}
\lab{eq:S-matrix}
\end{equation}
with $(x)_m\equiv\Gamma(x+m)/\Gamma(x)$, $3\le k \le N$ and $0\le m \le N-k$.
Similar relations hold in the anti-holomorphic sector, 
\begin{equation}
\wbar S_{11}=\e^{i\pi(N\wbar s-\wbar h-1)}\,,
\; \wbar S_{22}=\e^{i\pi(N\wbar s+\wbar
h-2)}\,,\;
\wbar S_{k,k+m}=(-1)^{k-3}\frac{(k-2\wbar s-1)_m}{m!}\,.
\end{equation}
The $S-$matrix does not depend on $z$
because
the $Q-$functions on the both sides of  relation  
(\ref{eq:S-def}) satisfy the
same differential equation (\ref{eq:Eq-1}).

Now, substituting (\ref{eq:Q-1}) and 
(\ref{eq:S-def}) into (\ref{eq:Qz-symmetry}), we find
\\[-1mm]
\begin{eqnarray}
\beta_h(q,\wbar q)&=& \e^{i\theta_N(q,\wbar q)} (-1)^{Nn_s+n_h}\beta_h(-q,-\wbar q)\,,\qquad
\nonumber
\\[1mm]
\gamma_{m\wbar m}(q,\wbar q) &=& \e^{i\theta_N(q,\wbar q)}
\sum_{n,\wbar n\ge 3}^N S_{nm} \gamma_{n\wbar n}(-q,-\wbar q)\,\wbar S_{\wbar n\wbar
m}\,.
\lab{eq:parity-qc}
\end{eqnarray}
In this way, similarly to the energy, Eq.~(\ref{eq:E-fin}),
which was calculated from the mixing matrix at $z=0$, the
eigenvalues of the quasimomentum, $\theta_N(q,\wbar q)$, 
can be calculated from
the mixing matrix at $z=1$, 
from the first relation in
(\ref{eq:parity-qc}). 
In the special case when
$q_{2k+1}=\wbar q_{2k+1}=0$
$(k=1,2\ldots )$, this means $\beta_h(q,\wbar q)=\beta_h(-q,-\wbar q)$, the 
quasimomentum
is equal to
\begin{equation}
\e^{i\theta_N(q,\wbar q)}=(-1)^{Nn_s+n_h}\,.
\lab{eq:quasi-0}
\end{equation}

\subsection{Transition matrices}

In the previous Sections we constructed
the solutions $Q(z,\wbar z)$ to (\ref{eq:Eq-1}) in
the vicinity of $z=0$ and $z=1$.
Now, we glue these solutions inside the region
$|1-z|<1,\,|z| < 1$ and, then analytically continue the
resulting expression for $Q(z,\wbar z)$ into the whole complex $z-$plane 
by making use of the duality relation (\ref{eq:Qz-symmetry}). 

Firstly, we define
the transition matrices $\Omega(q)$ and $\wbar \Omega(\wbar q)$:
\begin{equation}
Q_n^{(0)}(z)=\sum_{m=1}^N \Omega_{nm}(q)\, Q_m^{(1)}(z)\,,\qquad
\wbar Q_{n}^{(0)}(\wbar z)=\sum_{m=1}^N\wbar \Omega_{nm}(\wbar q)
 \,\wbar Q_{m}^{(1)}(\wbar z)\,.
\lab{eq:Omega-def}
\end{equation}
which are uniquely fixed \ci{Derkachov:2002wz}.
The resulting expressions for the matrices
$\Omega(q)$ and $\wbar\Omega(\wbar q)$ take the form of infinite series in $q$
and $\wbar q$, respectively. 
Substituting 
(\ref{eq:Omega-def}) into
(\ref{eq:Q-0}) and matching the result into 
(\ref{eq:Q-1}), we find 
the following
relation
\begin{equation}
C^{(1)}(q,\wbar q)=\left[\Omega(q)\right]^T C^{(0)}(q,\wbar q)\ \wbar
\Omega(\wbar{q})\,.
\lab{eq:C1-C0}
\end{equation}
The above matrix equation 
allows us to determine the matrices $C^{(0)}$ and $C^{(1)}$
and 
provides the quantization conditions for the integrals of
motion, $q_k$ and $\wbar q_k$ with $k=3,\ldots ,N$. 
Therefore, we can evaluate 
the eigenvalues of the Baxter $\mathbb{Q}-$operator,
Eq.~(\ref{eq:Q-R}). 
Formula (\ref{eq:C1-C0}) contains
$N^2$ equations with:
\begin{itemize}
\item $(N-1)$ $\alpha-$parameters inside the matrix $C^{(0)}$,
\item $2+(N-2)^2$ parameters $\beta_{1,2}$ 
and $\gamma_{m\wbar m}$ inside the matrix $C^{(1)}$,
\item $(N-2)$ integrals of motion $q_3,\ldots ,q_N$ where $\wbar
q_k=q_k^*$
\end{itemize}
Thus, we obtain
$(2N-3)$ nontrivial consistency conditions.

The solutions to the quantization conditions (\ref{eq:C1-C0}) 
will be presented in detail in 
next Sections.

\section{Properties of the $N-$Reggeon states}

In this Section we characterize the spectra of the conformal charges
obtained by numerical calculations \ci{Derkachov:2002wz,Kotanski:2001iq}.
Here, parametrization of the spectra is presented
and the spectrum symmetries are shown.
Moreover, descendent states are described.


\subsection{Trajectories}
Solving quantization conditions (\ref{eq:C1-C0})
we obtain continuous trajectories
in the space of conformal charges.
They are built of points, $(q_2(\nu_h),\ldots,q_N(\nu_h))$ which satisfy
(\ref{eq:C1-C0}) and depend on a continuous 
real parameter  $\nu_h$ entering $q_2$, (\ref{eq:q2}) and 
(\ref{eq:hpar}).
In order to label the trajectories 
we introduce the set of the integers 
\begin{equation}
\mybf{\ell}=\{\ell_1,\ell_2,\ldots,\ell_{2(N-2)}\}
\lab{eq:dell}
\end{equation}
which parameterize one specified point on each trajectory 
for given $h$.
Specific examples in the following sections will further clarify
this point. 

Next we calculate the observables
along these trajectories, 
namely the energy (\ref{eq:E-fin}) 
and the quasimomentum (\ref{eq:parity-qc}). 
The quasimomentum is constant (\ref{eq:quask}) for all points 
situated on a given trajectory.
The minimum of the energy, which means the maximal intercept,
for almost all trajectories is located
at $\nu_h=0$.
It turns out that the energy behaves around $\nu_h=0$ like
\begin{equation}
E_N(\nu_h;\mybf{\ell}^{\rm ground})
=E_N^{\rm ground}+\sigma_N {\nu_h}^2+{\cal O}({\nu_h}^2)
\lab{eq:Enu}
\end{equation}
Thus, the ground state along its trajectory is gapless 
and the leading contribution
to the scattering amplitude around $\nu_h$ may be rewritten as
a series in the strong coupling constant:
\begin{equation}
{\cal A}(s,t) \sim -i s 
\sum _{N=2}^\infty (i \asbar)^N
\frac{s^{-\asbar E_N^{\rm ground}/4}}{(\asbar\sigma_N
\ln s)^{1/2}}\,\xi_{A,N}(t)
\xi_{B,N}(t)\,,
\lab{eq:amp}
\end{equation}
where  $\asbar=\alpha_s N_c/\pi$ 
and $\xi_{X,N}(t)$ are the impact factors
corresponding to the overlap between the wave-functions of scattered
particle with the wave-function of $N-$Reggeons, whereas 
$\sigma_N$ measures the dispersion of the energy on the
the trajectory around $\nu_h=0$.

On the other hand,
the energy along the trajectories 
grows with $\nu_h$ and for $|\nu_h| \rightarrow \infty$ 
and finally, we have 
$E_N(\nu_h;{\mybf \ell}) \sim \ln {\nu_h}^2$. 
These parts of the trajectory give the lowest contribution to the
scattering amplitude.

\subsection{Symmetries}
The spectrum of quantized charges $q_2,\ldots,q_N$ is degenerate.
This degeneration is caused by two symmetries:
\begin{equation}
q_k \leftrightarrow (-1)^k q_k
\lab{eq:qkmsym}
\end{equation}
which comes from
invariance of the Hamiltonian under mirror permutations of particles, 
(\ref{eq:PMsym}), and
\begin{equation}
q_k \leftrightarrow  \wbar q_k
\lab{eq:qkcsym}
\end{equation}
which is 
connected with
the symmetry under interchange of the $z-$ and 
$\wbar z-$sectors.
Therefore, the four points, $\{q_k\}$, $\{(-1)^k q_k\}$,
$\{{q_k}^\ast\}$ and $\{(-1)^k {q_k}^{\ast}\}$ with $k=2,\ldots,N$, 
are related and all of them satisfy the quantization conditions 
(\ref{eq:C1-C0})
and have the same energy.

\subsection{Descendent states}

Let us first discuss
the spectrum along the trajectories
with the highest conformal charge $q_N$
equal zero
for arbitrary $\nu_h \in \mathbb{R}$.
It turns out \ci{Vacca:2000bk,Lipatov:1998as,Bartels:1999yt,Derkachov:2002wz} 
that the wave-functions of these states
are built of $(N-1)-$particle states. Moreover, their energies
\ci{Korchemsky:1994um} are also equal to the energy of the ancestor
$(N-1)-$particle states: 
\begin{equation}
E_N(q_2,q_3,\ldots,q_N=0)=E_{N-1}(q_2,q_3,\ldots,q_{N-1})\,.
\lab{eq:Edes}
\end{equation}
Thus, we call them the descendent states
of the $(N-1)-$particle states.

Generally, for odd $N$, the descendent state $\Psi_N^{(q_N=0)}$
with the minimal energy $E_N(q_N=0)=0$ 
has 
for $q_2=0$, \ie for $h=0,1$,  the remaining 
integrals of motion 
$q_3=\ldots =q_N=0$ as well.
For $h=1+i \nu_h$, \ie $q_2 \ne 0$,
the odd conformal charges $q_{2k+1}=0$ with $k=1,\ldots,(N-1)/2$
while the even ones $q_{2k}\ne 0$ and depend on $\nu_h$.

On the other hand, for even $N$, the eigenstate with the minimal energy 
$\Psi_N^{(q_N=0)}$  is the descendent state of the $(N-1)-$particle 
state 
which has minimal energy with $q_{N-1} \ne 0$.  Thus, 
${E_N^{min}(q_N=0)=E_{N-1}^{min}(q_{N-1}\ne0)>0}$.

Studying more exactly this problem one can obtain
\ci{Vacca:2000bk,Derkachov:2002wz}
a relation between the quasimomenum $\theta_N$ of the descendent state 
and the ancestor one  $\theta_{N-1}$, which takes the following form
\begin{equation}
\e^{i \theta_N} \bigg|_{q_{N}=0}=
-\e^{i \theta_{N-1}}=
(-1)^{N+1}\,.
\lab{eq:quasdes}
\end{equation} 

Additionally, one can define 
a linear operator $\Delta$  
\ci{Vacca:2000bk,Derkachov:2002wz}
that maps 
the subspace $V_{N-1}^{(q_N-1)}$ of the $(N-1)-$particle 
ancestor eigenstates 
with the quasimomentum $\theta_{N-1}=\pi N$
into the $N-$particle descendent states with $q_N=0$
and $\theta_N=\pi (N+1)$ as
\begin{equation}
\Delta:
\quad
V_{N-1}^{(\theta_{N-1}=\pi N)} \to 
V_{N}^{(\theta_N=\pi (N+1))} \;.
\lab{eq:Delta}
\end{equation}
It turns out that this operator is nilpotent 
for the eigenstates 
which form trajectories
\ci{Vacca:2000bk}, \ie $\Delta^2 \Psi=0$.
Thus, the descendent-state trajectory 
can not be ancestor one for $(N+1)-$particle states.
However, it is possible to built a single  state \ci{Derkachov:2002wz}
with $q_2=q_3=\ldots=q_N=0$, \ie for only one point $\nu_h=0$,
that has $E_N=0$ and the eigenvalue of Baxter $Q-$operator
defined as
\begin{equation}
Q_N^{q=0}(u,\wbar u) \sim \frac{u-\wbar u}{\wbar u^N}\,,
\lab{eq:Qq0}
\end{equation}
where a normalization factor was omitted.

Additional 
examples of the descendent states will be 
described later in the next sections.

\section{Quantum numbers of the $N=4$ states}

In this Section we present the spectrum for four Reggeons
calculated by making use of $Q-$Baxter method. To this end we 
resum solutions (\ref{eq:power-series-0}) and 
(\ref{eq:v-series}) numerically, and using them we solve the quantization
conditions (\ref{eq:C1-C0}). 
Earlier, some results for $N=4$ were only presented 
in \ci{Derkachov:2002wz} and some numerical results in Ref.
\ci{deVega:2002im}.
Here we show much more data and we present
more detailed analysis of this spectrum. i.e.
resemblant and winding
spectra of $q_3$, $q_4$  and
corrections to the WKB approximation for $N=4$.

For four Reggeons the spectrum of the conformal charges 
is much more complicated than in the three-Reggeon case 
\ci{Kotanski:2006ec}.
Indeed,
we have here the space of three conformal charges $(q_2,q_3,q_4)$.
Thus, apart from the lattice structure in $q_4^{1/4}$ 
we have also respective lattice structures
in $q_3-$space.
Here we consider the case for $n_h=0$ so that $h=\frac{1}{2}+i \nu_h$. 
This spectrum includes
the ground states.
For clarity we split these spectra into several parts. 
We perform this separation by 
considering spectra with different quasimomenta 
$\theta_4(q,\wbar q)$ as well as
a different quantum number $\ell_3$,
which will be defined in the solution (\ref{eq:theta-4})--(\ref{eq:q3-quan})
of two quantization conditions (\ref{eq:C1-C0}) for $N=4$ 
from Ref. \ci{Derkachov:2002pb}.


From the first quantization condition 
of WKB approximation \ci{Derkachov:2002pb}
one can get the WKB approximation of
the charge $q_4$ as
\begin{equation}
q_4^{1/4}=\frac{\Gamma^2(3/4)}{4\sqrt{\pi}}\left[ \frac1{\sqrt {2}}
\ell_1+
\frac{i}{\sqrt {2}}\ell_2 \right]
\lab{eq:q4-quan}
\end{equation}
and the quasimomentum
 is equal to
\begin{equation}
\theta_4=-\frac{\pi}2\ell
=\frac{\pi}2 
(\ell_2+\ell_3-\ell_1)\qquad ({\rm mod}~ 2\pi)\,,
\lab{eq:theta-4}
\end{equation}
where $\ell_1$, $\ell_2$ and $\ell_3$ are even for even $\ell$ and
odd for odd $\ell$.
Thus, we have two kinds of lattices: with $\theta_4=0,\pi$ and
with $\theta_4=\pm \pi/2$. 
They are presented in Fig.\ \ref{fig:Q4}. In these pictures
gray
lines show the WKB lattice (\ref{eq:q4-quan})
with vertices at $\ell_1, \ell_2 \in \mathbb{Z}$.
To find  the leading approximation 
for the charge $q_3$, we apply the second relation of WKB
approximation \ci{Derkachov:2002pb} gives
\begin{equation}
\Im \frac{q_3}{q_4^{1/2}} = (\ell_1-\ell_2-\ell)=\ell_3.
\lab{eq:q3-quan}
\end{equation}
Notice that the system (\ref{eq:q4-quan}) and (\ref{eq:q3-quan}) 
is underdetermined and 
it does not fix the charge $q_3$ completely \ci{Derkachov:2002pb}.

\medskip
\begin{figure}[!ht]
\centerline{
\begin{picture}(165,60)
\put(20,2){\epsfysize6.0cm \epsfbox{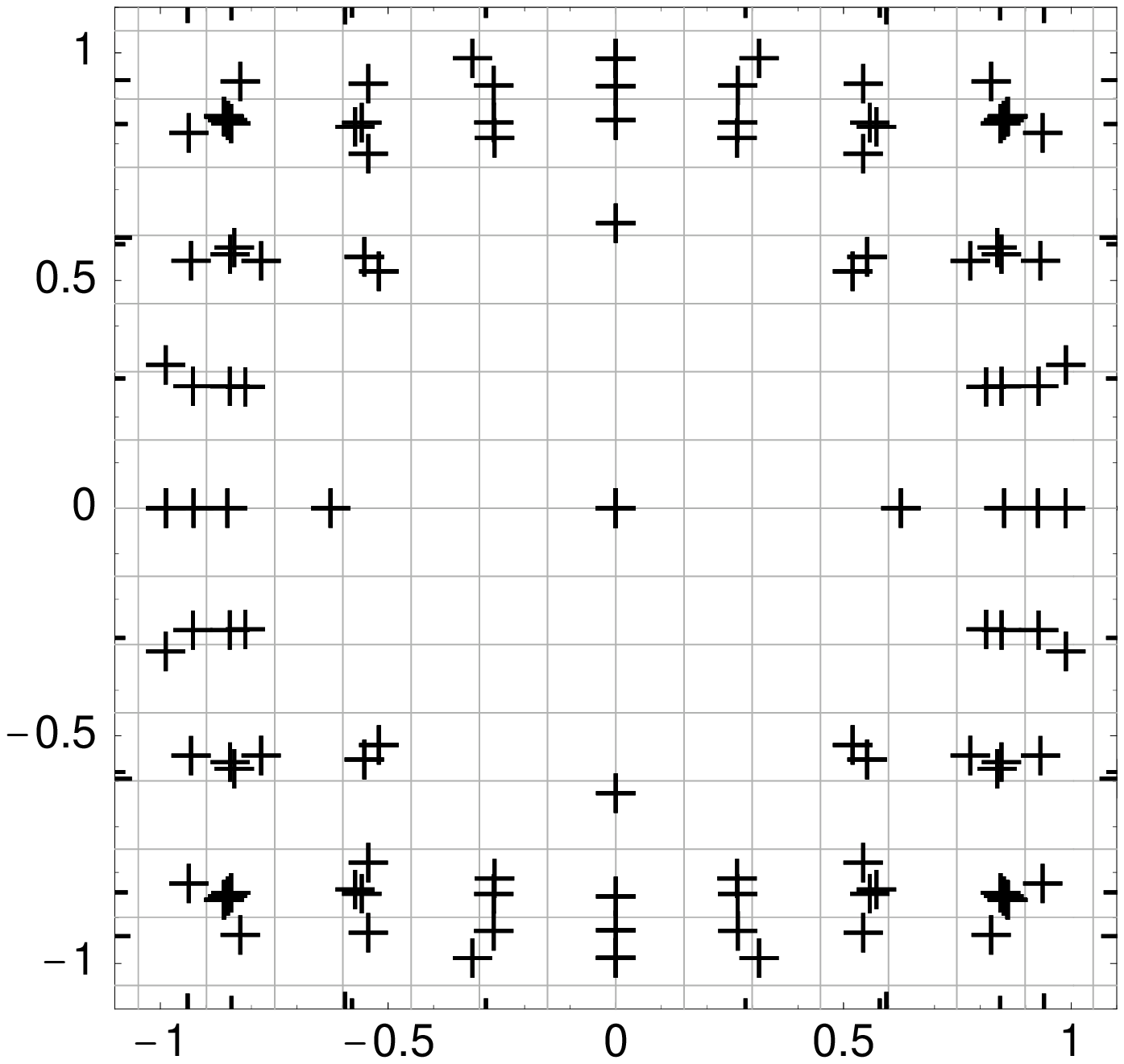}}
\put(85,2){\epsfysize6.0cm \epsfbox{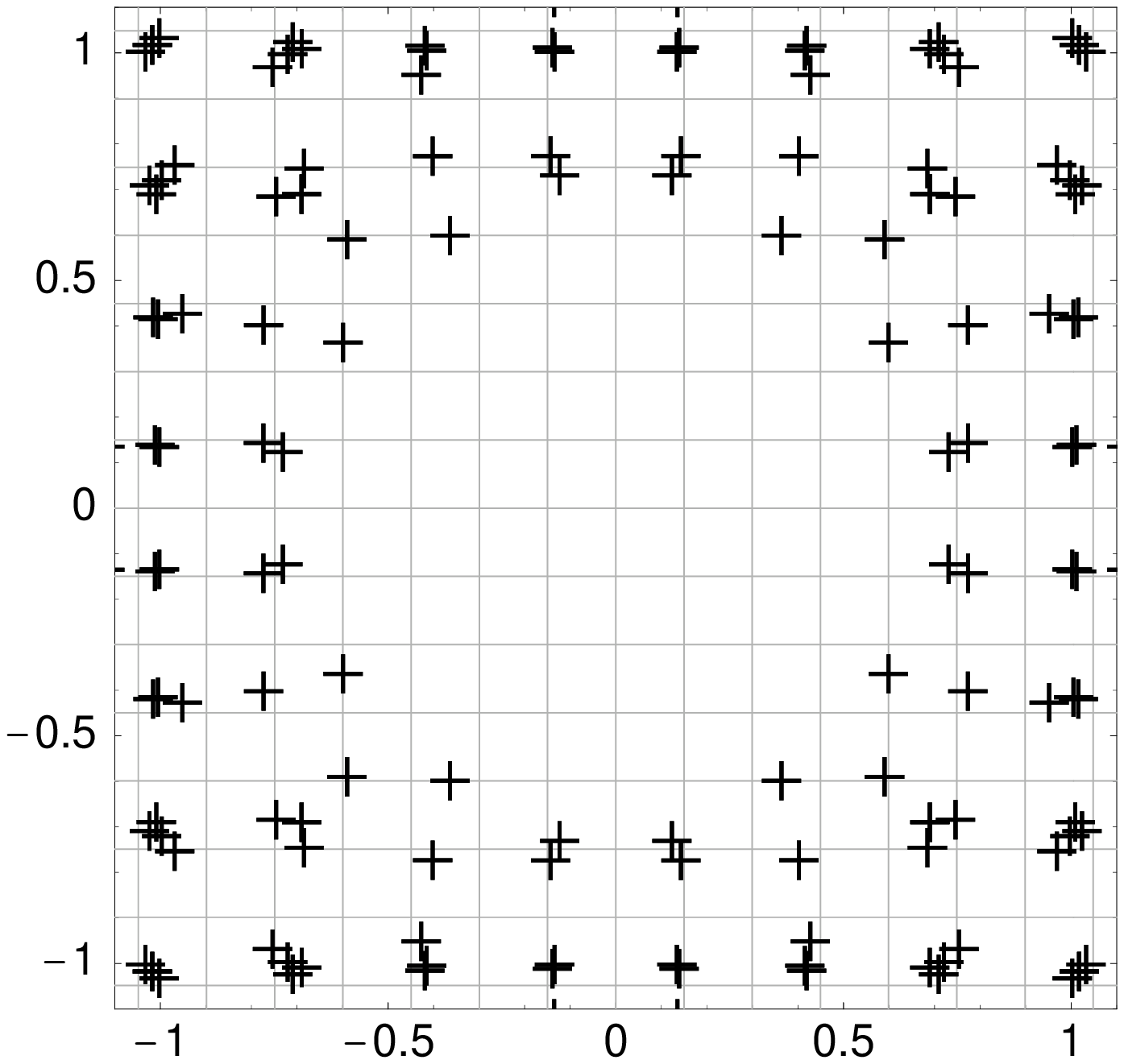}}
\put(48,0){$\RRe[q_4^{1/4}]$}
\put(18,27){\rotatebox{90}{$\IIm[q_4^{1/4}]$}}
\put(113,0){$\RRe[q_4^{1/4}]$}
\put(83,27){\rotatebox{90}{$\IIm[q_4^{1/4}]$}}
\end{picture}
}
\caption[The spectrum of  $q_4$ for $N=4$ 
and the total spin $h=1/2$.]{The spectrum of the integrals of 
motion $q_4$ for $N=4$ and the total spin $h=1/2$.
The left and right panels correspond to the eigenstates with different
quasimomenta $\e^{i\theta_4}=\pm 1$ and $\pm i$, respectively.}
\lab{fig:Q4}
\end{figure}

It turns out that after choosing one value of $\theta_4$,
the lattice in $q_4^{1/4}-$space 
is still spuriously degenerated\footnote{degeneration in the 
leading order of the WKB approximation} and this degeneration 
also corresponds to different lattices
in $q_3^{1/2}$. The parameter $\ell_3$ which is defined in (\ref{eq:q3-quan})
will be used to distinguish these different lattices.

\subsection{Descendent states for $N=4$}

One can notice that for $N=4$ and $n_h=0$ we have the descendent states.
They appear in sector
with the quasimomentum $\theta_4=\pi$, which agrees with (\ref{eq:quasdes}).
The wave-functions of the descendent states
are built of three-particle eigenstates  with $\theta_3=0$.
Additionally, the spectrum of $q_3$ for
these three-Reggeon states 
and the descendent state for $N=4$ 
is the same and it is depicted in 
Fig.\ \ref{fig:WKB-N4} on the left panel.
Moreover, the energy of these descendent state and
the corresponding three-Reggeon states are also the same (\ref{eq:Edes}).

\medskip
\begin{figure}[ht!]
\centerline{
\begin{picture}(165,60)
\put(20,2){\epsfysize6.0cm \epsfbox{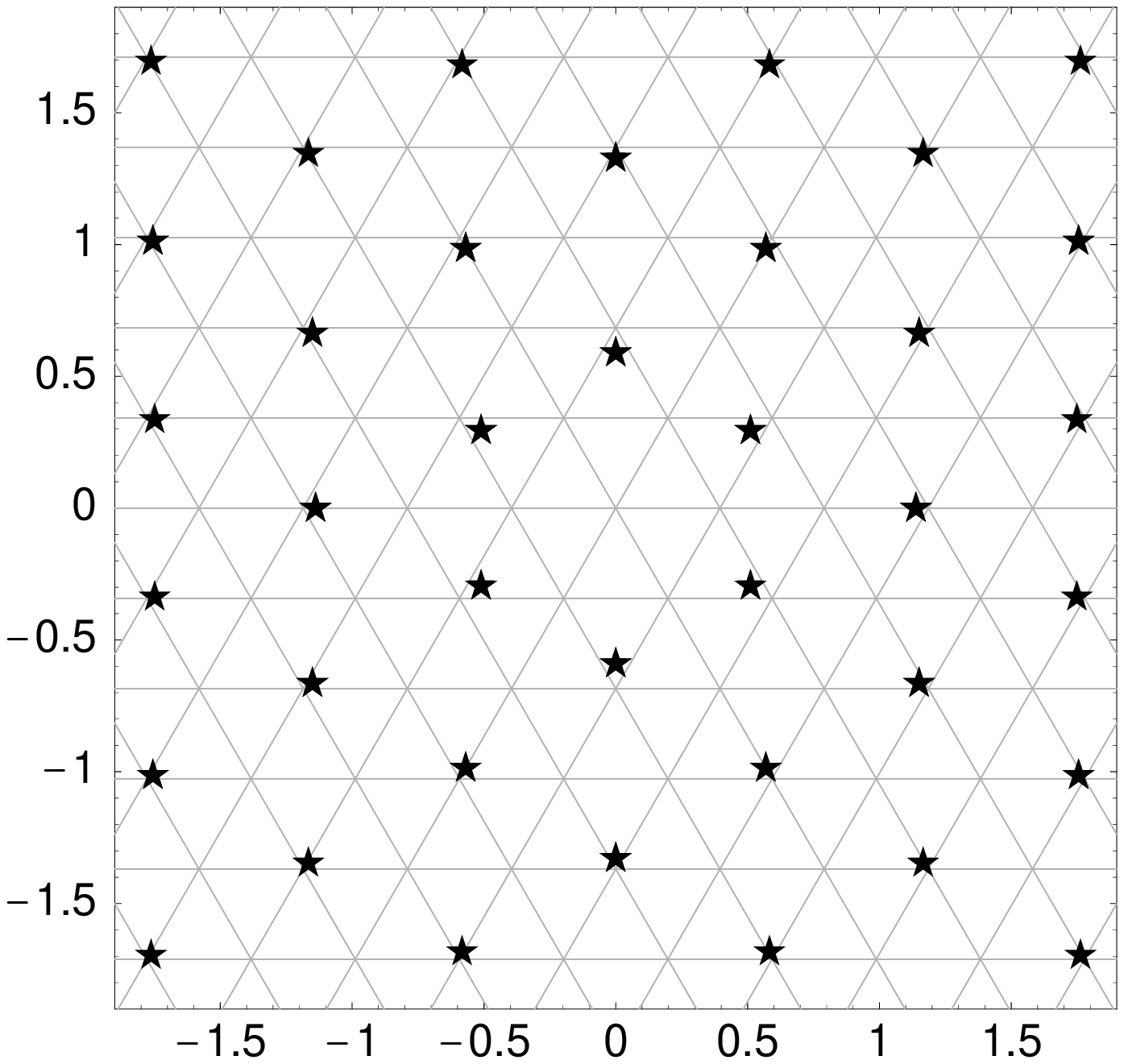}}
\put(85,2){\epsfysize6.0cm \epsfbox{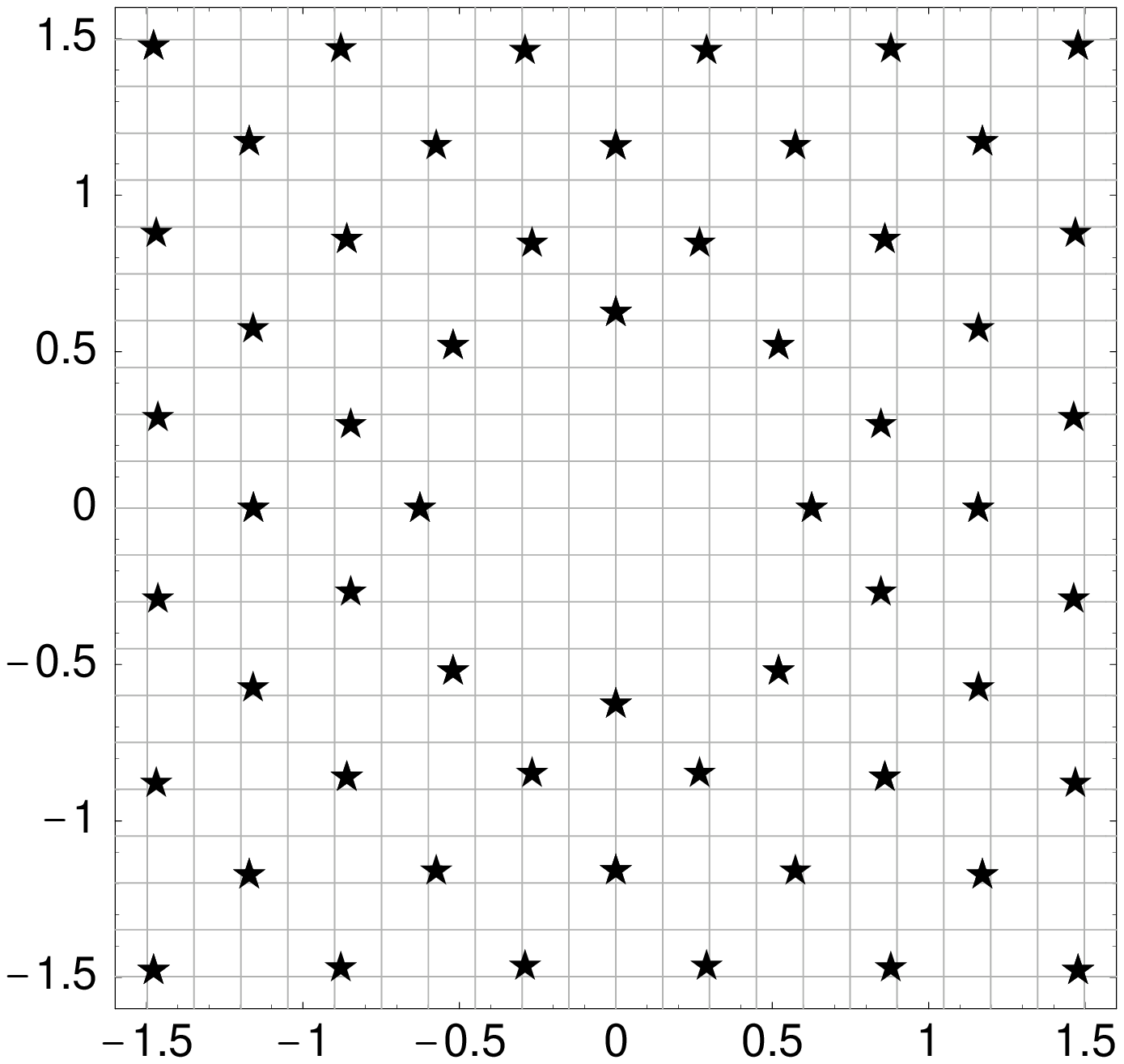}}
\put(48,0){$\RRe[q_3^{1/3}]$}
\put(18,27){\rotatebox{90}{$\IIm[q_3^{1/3}]$}}
\put(113,0){$\RRe[q_4^{1/4}]$}
\put(83,27){\rotatebox{90}{$\IIm[q_4^{1/4}]$}}
\end{picture}
}
\caption[The spectra of the conformal charges for $N=4$
and comparison to the WKB expansion.]{
The spectra of the conformal charges for $N=4$
and comparison to the WKB expansion. On the left panel, the spectrum
of $q_3$ with $q_4=0$ corresponding
to the descendent states with $\theta_4=\pi$. On the right panel, 
the spectrum of $q_4$ for $h=1/2$ and $q_3=0$ with $\theta_4=0$. 
The WKB lattices are denoted by the 
gray
lines.}
\lab{fig:WKB-N4}
\end{figure}

\subsection{Lattice structure for $q_3=0$}

Let us consider the spectrum with $q_4 \ne 0$ and 
$q_3=0$, 
see the right panel of Fig \ref{fig:WKB-N4}.
In this case the quasimomentum $\theta_4=0$ and 
the lattice structure
include vertices that correspond to the ground state.

Similarly to the $N=3$ case \ci{Kotanski:2006ec} 
we have in the $q_4^{1/4}-$space a lattice 
with a square-like structure described by (\ref{eq:q4-quan}). 
In this case even numbers $\ell_1$ and $\ell_2$ 
satisfy $\ell_1+\ell_2 \in 4 \mathbb{Z}$.
Thus, we have
the WKB formula
\begin{equation}
\left[q_4^{\rm WKB}(\ell_1,\ell_2)\right]^{1/4}
=\Delta_{N=4}\cdot\left(\frac{\ell_1}{2 \sqrt{2}}
+i\frac{\ell_2}{2 \sqrt{2}}\right),
\lab{eq:WKB-N4}
\end{equation}
where 
the vertices 
are placed outside a disk
around the origin of the radius
\begin{equation}
\Delta_{4}
=\left[\frac{4^{3/4}}{\pi}\int_{-1}^1\frac{dx}{\sqrt{1-x^4}}\right]^{-1}
=\frac{\Gamma^2(3/4)}{2\sqrt{\pi}}=0.423606\ldots \, .
\lab{eq:D4}
\end{equation}
As before, the leading-order WKB formula (\ref{eq:WKB-N4}) is valid only for
$|q_4^{1/4}|\gg |q_2^{1/2}|$. 

\begin{figure}[ht!]
\vspace*{5mm}
\centerline{
\psfrag{-E_4/4}[cc][cc]{$-E_4/4$}
\psfrag{nu_h}[cc][bc]{$\mbox{\large$\nu_h$}$}
\psfrag{q4/q2}[cc][cc]{$\, q_{_{\scriptstyle 4}}/q_{_{\scriptstyle 2}}$}
{\epsfysize4.4cm \epsfbox{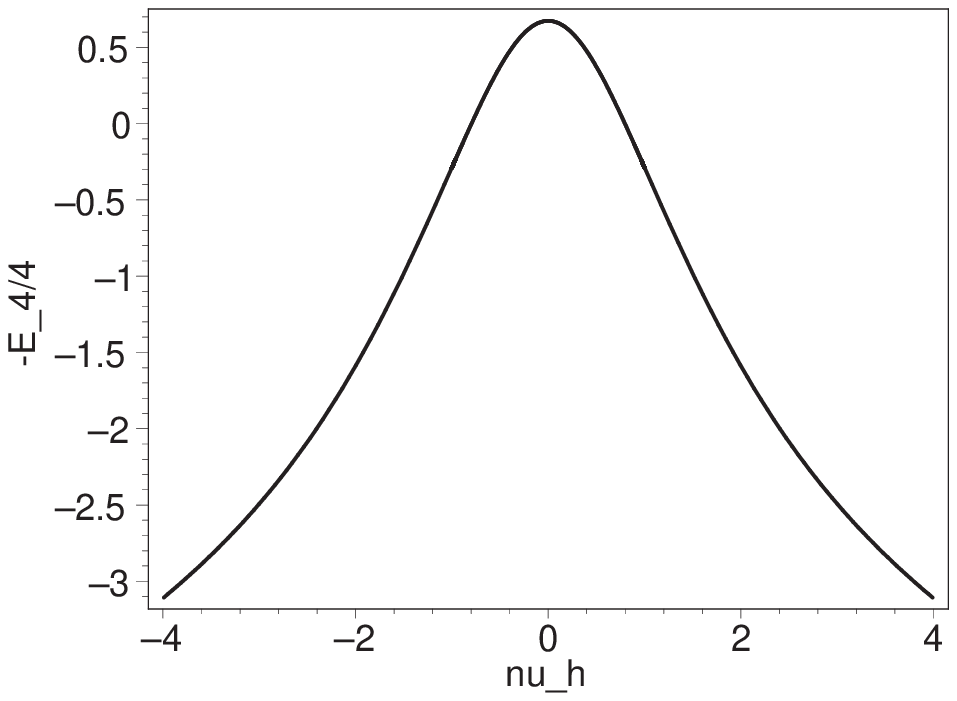}}\quad{\epsfysize4.4cm 
\epsfbox{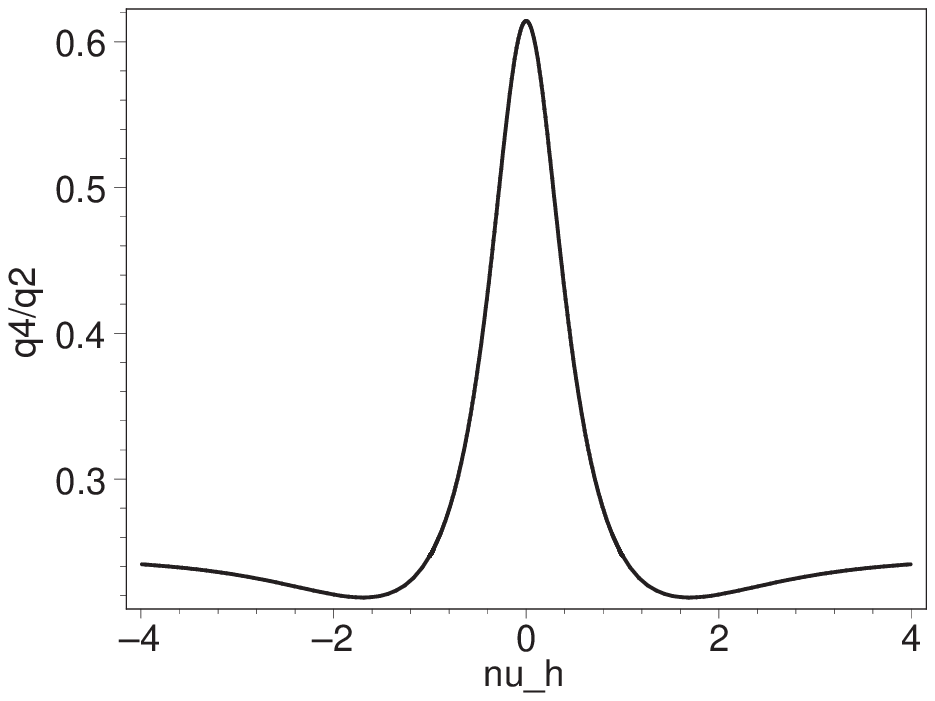}}}
\caption[The dependence of the energy, $-E_4/4$, and the quantum number,
 $q_4/q_2$,
with $q_2=1/4+\nu_h^2$]
{The dependence of the energy, $-E_4/4$, and the quantum number,
 $q_4/q_2$,
with $q_2=1/4+\nu_h^2$, on the total spin $h=1/2+i\nu_h$ along the ground state
trajectory for $N=4$.}
\lab{fig:N4-flow}
\end{figure}

The energy is lower for points  which are nearer to the origin. 
Similarly to the $N=3$ case, the spectrum is also built of trajectories which
extend  in the $(\nu_h,q_3,q_4)-$space.
Namely, each
point on the $q_4^{1/4}-$lattice belongs to one specific 
trajectory parameterized by
the set of integers $\{\ell_1,\ell_2,\ldots\}$. 

The ground state for $N=4$ is situated on a trajectory with
$(\ell_1,\ell_2)=(4,0)$.
We find that for this trajectory 
$q_3=\Im q_4=0$, 
whereas $\Re[q_4]$ and $E_4$ vary with $\nu_h$ as we show in
Figure~\ref{fig:N4-flow}. 
An accumulation of the energy levels 
in the vicinity of $\nu_h=0$ is described by Eq.~(\ref{eq:Enu}) with 
the dispersion parameter $\sigma_4=5.272$.

On the $q_4^{1/4}-$plane the ground state is 
represented by four points with the
coordinates $(\ell_1,\ell_2)=(\pm 4,0)$ and $(0,\pm 4)$.
Due to a residual
symmetry $q_4^{1/4} \leftrightarrow \exp(ik \pi/2) q_4^{1/4}$, 
they describe a single eigenstate with  
\begin{equation}
q_3^{\rm ground}=0\,,\qquad q_4^{\rm ground}=0.153589\ldots\,,
\qquad E_4^{\rm ground}=-2.696640\ldots\,.
\end{equation}
with $h=1/2$.  
It has the quasimomentum $\theta_4=0$ and, in contrast to the
$N=3$ case, it is unique.

\begin{table}[ht]
\begin{center}
\begin{tabular}{|c||c|c|c|c|c|c|}
\hline
 $(\ell_1/2,\ell_2/2)$&  $\left(q_4^{\rm \,exact}\right)^{1/4}$ & 
$\left(q_4^{\rm WKB}\right)^{1/4}$ & $-E_4/4$ \\
\hline
\hline
$(2,0)$ & $0.626 $ &  $0.599 $ & $0.6742$ \\ \hline
$(2,2)$ & $0.520+0.520\,i$ & $0.599+0.599\,i$ & $-1.3783$ \\ \hline
$(3,1)$ & $0.847+0.268\,i$ &  $0.899+0.299\,i$  & $-1.7919$ \\ \hline 
$(4,0)$ & $1.158$  &   $1.198$ & $-2.8356$ \\ \hline
$(3,3)$ & $0.860+0.860\,i$ & $0.899+0.899\,i$  & $-3.1410$  \\ \hline
$(4,2)$ & $1.159+0.574\,i$ & $1.198+0.599\,i$ & $-3.3487$ \\ \hline
\end{tabular}
\end{center}
\caption{Comparison of the exact spectrum of $q_4^{1/4}$ at $q_3=0$ 
and $h=1/2$ with the approximate
WKB expression \ref{eq:WKB-N4}. The last column shows the exact energy $E_4$.}
\lab{tab:WKB-4}
\end{table}

The comparison of (\ref{eq:WKB-N4}) with the 
exact
results for $q_4$ at $h=1/2$ is shown in Figure~\ref{fig:WKB-N4} and
Table~\ref{tab:WKB-4}. One can see that 
the WKB formula (\ref{eq:WKB-N4}) 
describes the spectrum with a good accuracy.

\subsection{Resemblant lattices with $\ell_3=0$}

In the previous Section we introduced the parameter $\ell_3$
which helps us to distinguish different lattices. This parameter
takes even values for $\theta_4=0,\pi$ and odd ones for 
$\theta_4=\pm \pi/2$. Let us take $\ell_3=0$. It turns out that 
in this case the spectrum lattice in the $q_3^{1/2}-$space is similar
to the corresponding lattice in 
the $q_4^{1/4}-$space, \ie
considering only the 
leading order of the WKB approximation
the $q_3^{1/2}-$lattice in comparison
to the $q_4^{1/4}-$lattice 
is rescaled by some real number.
An example of such a lattice for
$\theta_4=0$
is shown
in Figure \ref{fig:l30k01}. 
One can notice that the non-leading corrections to the WKB approximation
cause the bending of the lattice structure in  Fig.\ \ref{fig:l30k01}: 
for $q_4^{1/4}$ concave whereas for $q_3^{1/2}$ convex. 
Moreover, we can see the $q_4^{1/4}-$lattice as well as
$q_3^{1/2}-$one have a similar structure to the lattice with $q_3=0$
presented in Fig.\ \ref{fig:WKB-N4}. 
These lattices also do not have 
the vertices inside the disk at the origin $q_3=q_4=0$.
\medskip
\begin{figure}[ht!]
\centerline{
\begin{picture}(165,60)
\put(21,2){\epsfysize6.0cm \epsfbox{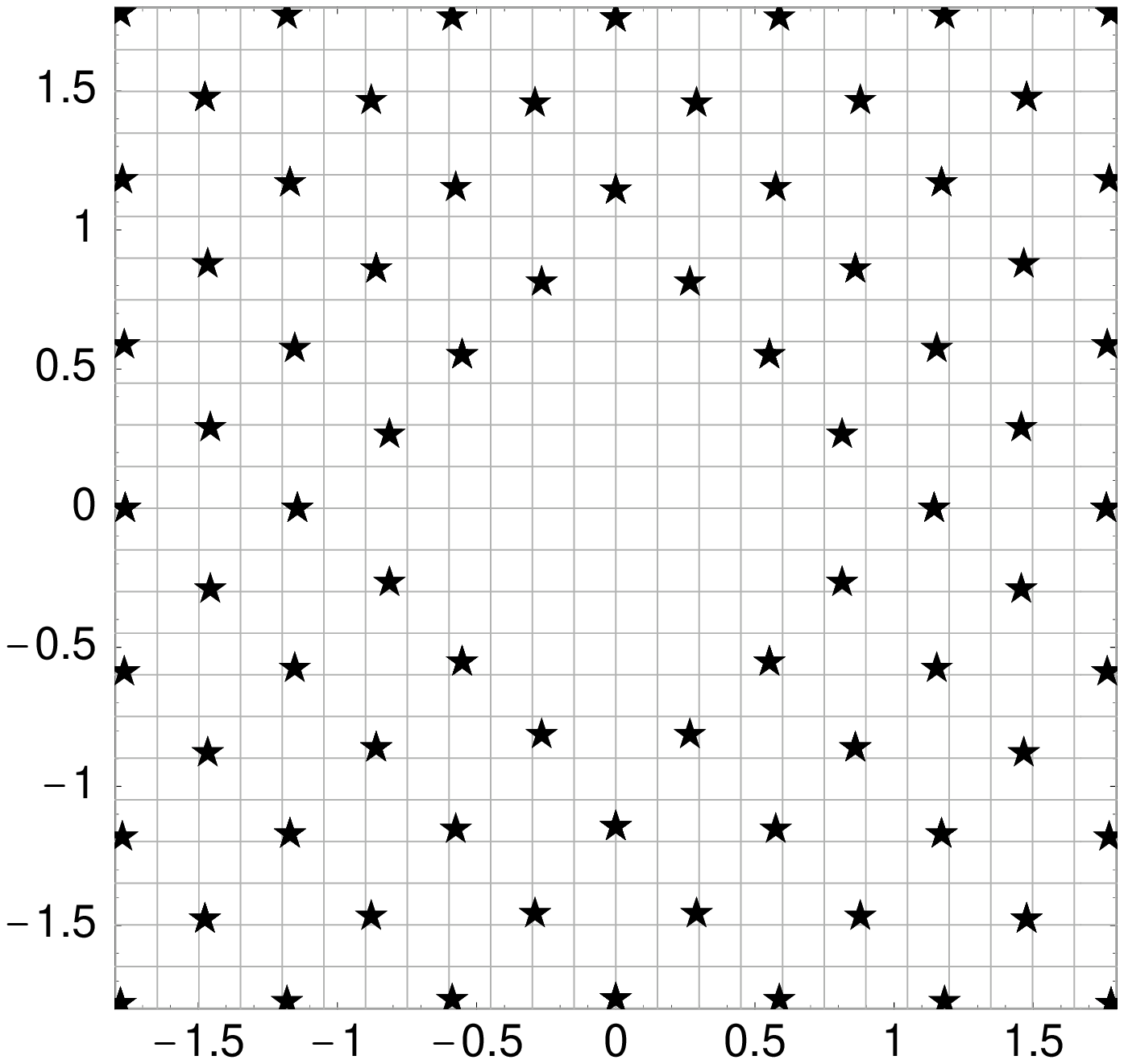}}
\put(85,2){\epsfysize6.0cm \epsfbox{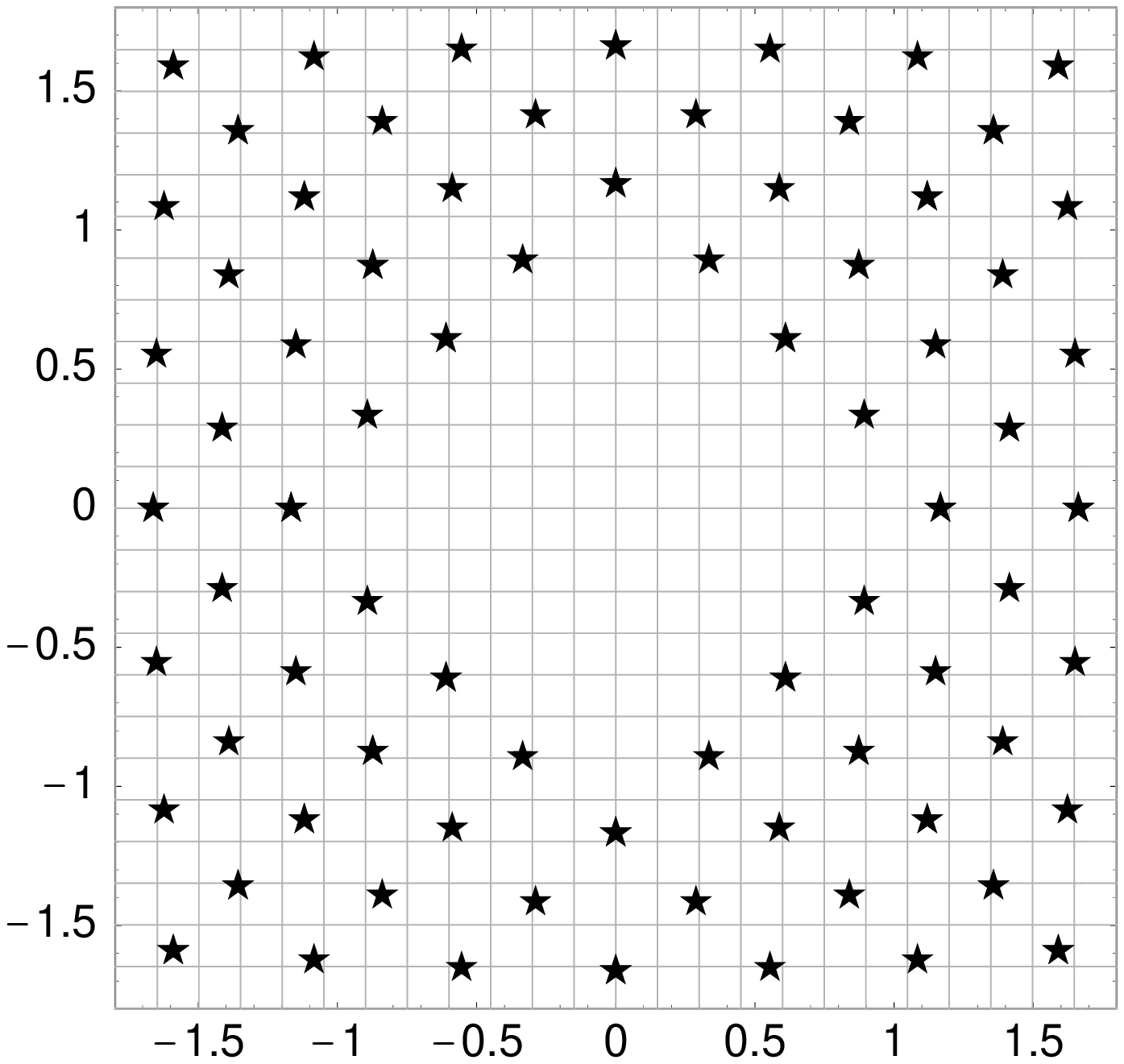}}
\put(48,0){$\RRe[q_3^{1/4}]$}
\put(18,27){\rotatebox{90}{$\IIm[q_3^{1/4}]$}}
\put(113,0){$\RRe[q_4^{1/2}]$}
\put(83,27){\rotatebox{90}{$\IIm[q_4^{1/2}]$}}
\end{picture}
}
\caption[The spectra of the conformal charges 
for $N=4$ with $\theta_4=0$,
$\ell_3=0$ and $\ell_4=1$.]{The  
spectra of the conformal charges for $N=4$ with $\theta_4=0$,
$\ell_3=0$ and $\ell_4=1$. On the left panel 
the spectrum of $q_4^{1/4}$, while on the right
panel the spectrum of $q_3^{1/2}$}
\lab{fig:l30k01}
\end{figure}


\medskip
\begin{figure}[ht!]
\centerline{
\begin{picture}(165,60)
\put(21,2){\epsfysize6.0cm \epsfbox{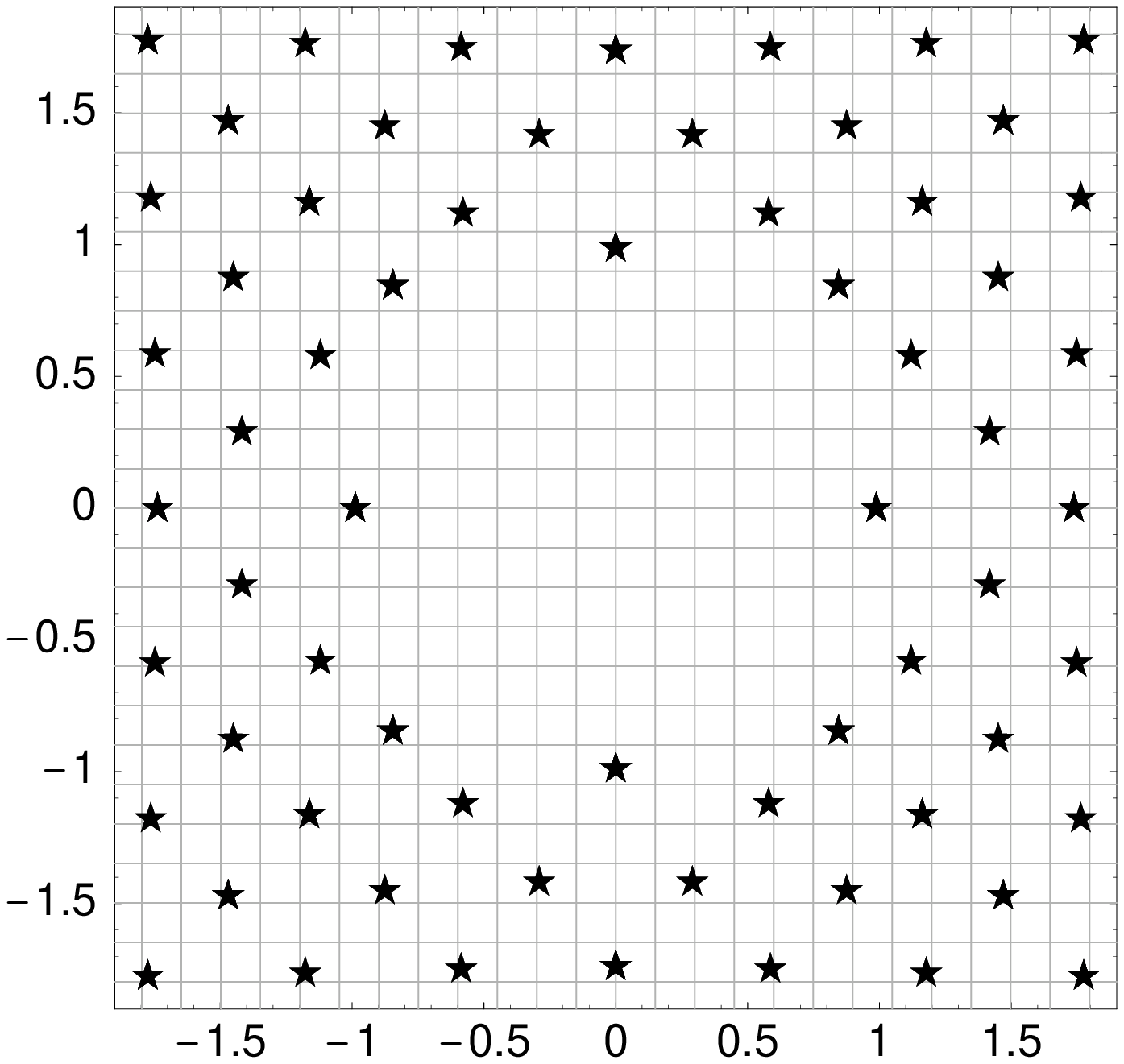}}
\put(87,3){\epsfysize5.8cm \epsfbox{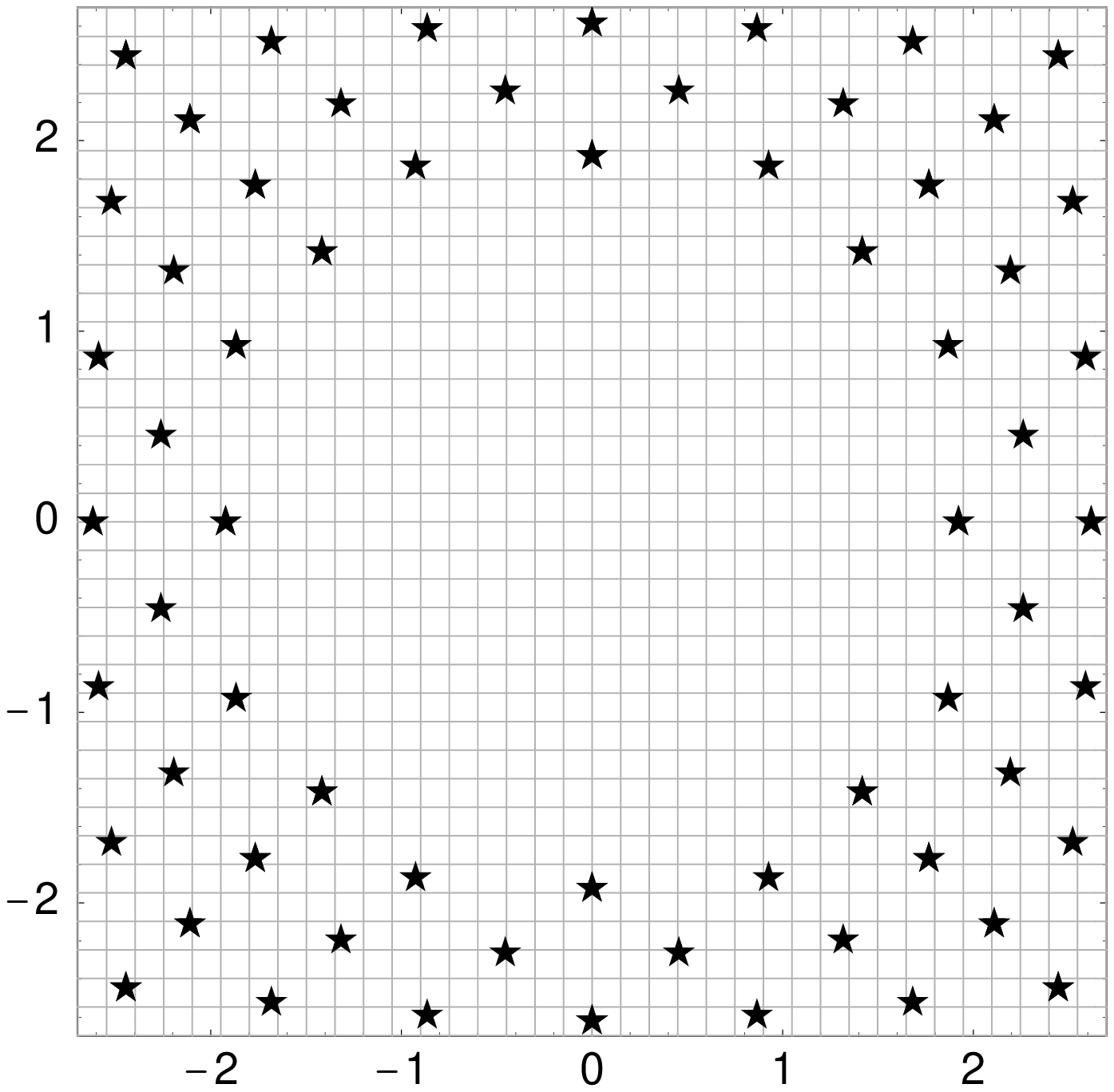}}
\put(48,0){$\RRe[q_3^{1/4}]$}
\put(18,27){\rotatebox{90}{$\IIm[q_3^{1/4}]$}}
\put(113,0){$\RRe[q_4^{1/2}]$}
\put(83,27){\rotatebox{90}{$\IIm[q_4^{1/2}]$}}
\end{picture}
}
\caption[The spectra of the conformal charges 
for $N=4$ with $\theta_4=0$,
$\ell_3=0$ and $\ell_4=2$.]{The spectra of 
the conformal charges for $N=4$ with 
$\theta_4=0$, $\ell_3=0$ and $\ell_4=2$. 
On the left panel the spectrum of $q_4^{1/4}$, while on the right
panel the spectrum of $q_3^{1/2}$}
\lab{fig:l30k02}
\end{figure}

Substituting 
\begin{equation}
q_3^{1/2}=r_3 \e^{i\phi_3}
\quad
\mbox{and}
\quad
q_4^{1/4}=r_4 \e^{i\phi_4}
\lab{eq:qrphi}
\end{equation}
into (\ref{eq:q3-quan})
we obtain a condition for the leading order of the WKB approximation
\begin{equation}
\ell_3=\left(\frac{r_3}{r_4}\right)^2 \sin \left(2 (\phi_3 -\phi_4) \right).
\lab{eq:q3-rphi}
\end{equation}
Thus, for $\ell_3=0$ and for a scale $\lambda=r_3/r_4 >0$ we have
$\phi_3=\phi_4$. This means that the vertices on the $q_3^{1/2}-$lattice
have the same angular coordinates as those from the $q_4^{1/4}-$lattice.
Looking at the numerical results in Figure \ref{fig:l30k01}
we notice that the missing quantization condition for $\ell_3=0$ in
the leading WKB order should 
have a form similar to
\begin{equation}
\left| \frac{q_3}{q_4^{1/2}} \right|={\lambda_{\ell_4}}^2\,,
\lab{eq:conlamb}
\end{equation}
where $\lambda_{\ell_4}=r_3/r_4 \in \mathbb{R}$ is a constant scale for
a given lattice $\ell_4$.

It turns out that
for a specified quasimomentum we have an infinite number of such lattices.
They differ from each other by a given scale $\lambda_{\ell_4}$.
For example for $\theta_4=0$ we have another lattice, shown 
in Fig \ref{fig:l30k02}. Its scale 
$\lambda_2$
differs from
the scale $\lambda_1$ of the lattice from Fig.\ \ref{fig:l30k01}.
The resemblant $q_4^{1/4}-$lattices for $\theta_4=0$ are described
by (\ref{eq:q4-quan}) with the integer parameters $\ell_1$ and $\ell_2$
satisfying $\ell_1+\ell_2 \in 4 \mathbb{Z}$.

The similar lattices also exist in the sector 
with the quasimomentum $\theta_4=\pi$.
Some of them are presented in Figures \ref{fig:l30k21}
and \ref{fig:l30k22}.
For $\theta_4=\pi$
the resemblant  $q_4^{1/4}-$lattices are also described 
by (\ref{eq:q4-quan}) but the integer parameters $\ell_1$ and $\ell_2$
satisfy $\ell_1+\ell_2 \in 4 \mathbb{Z}+2$.
\medskip
\begin{figure}[ht!]
\centerline{
\begin{picture}(165,60)
\put(21,2){\epsfysize6.0cm \epsfbox{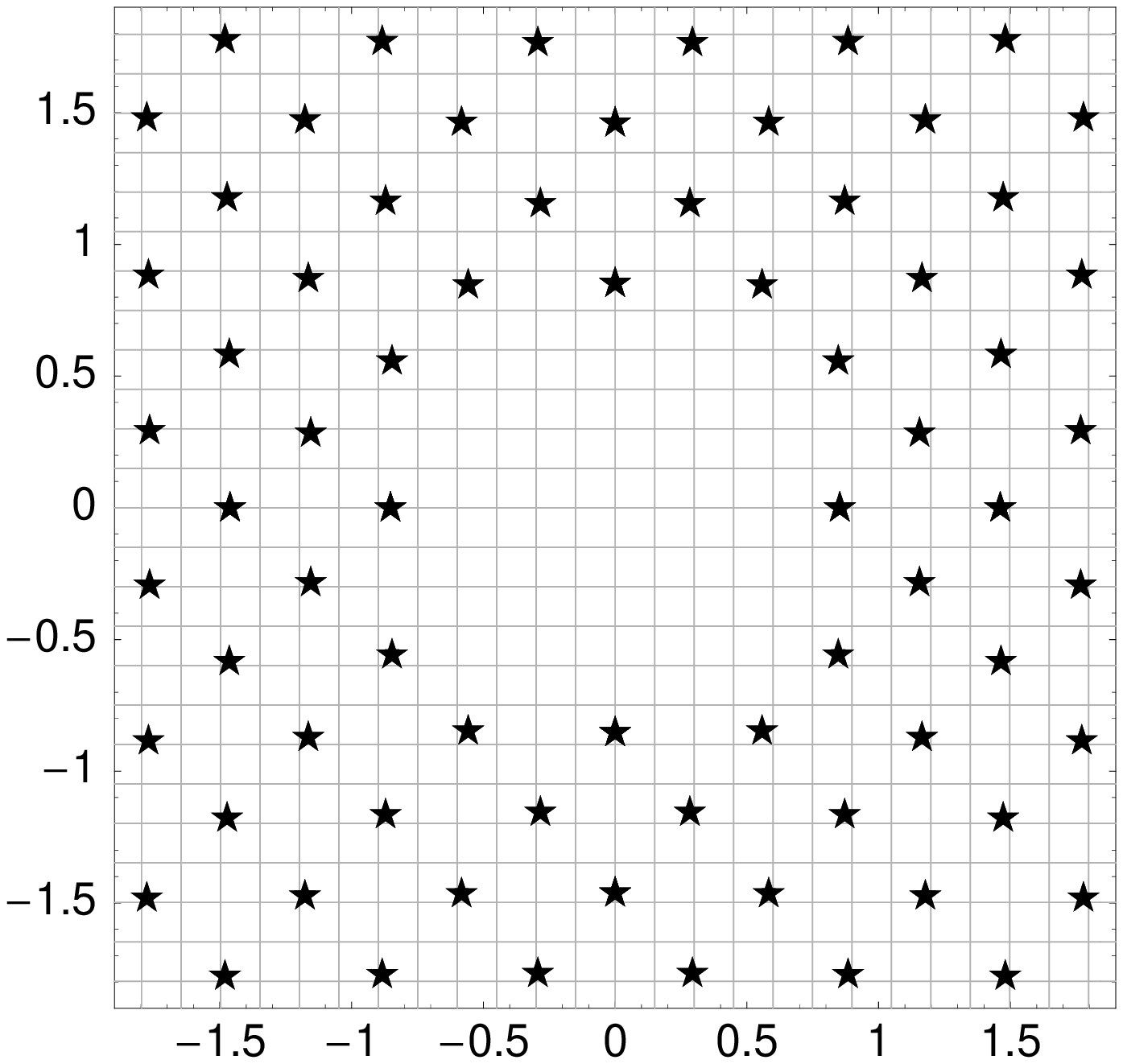}}
\put(85,2){\epsfysize6.0cm \epsfbox{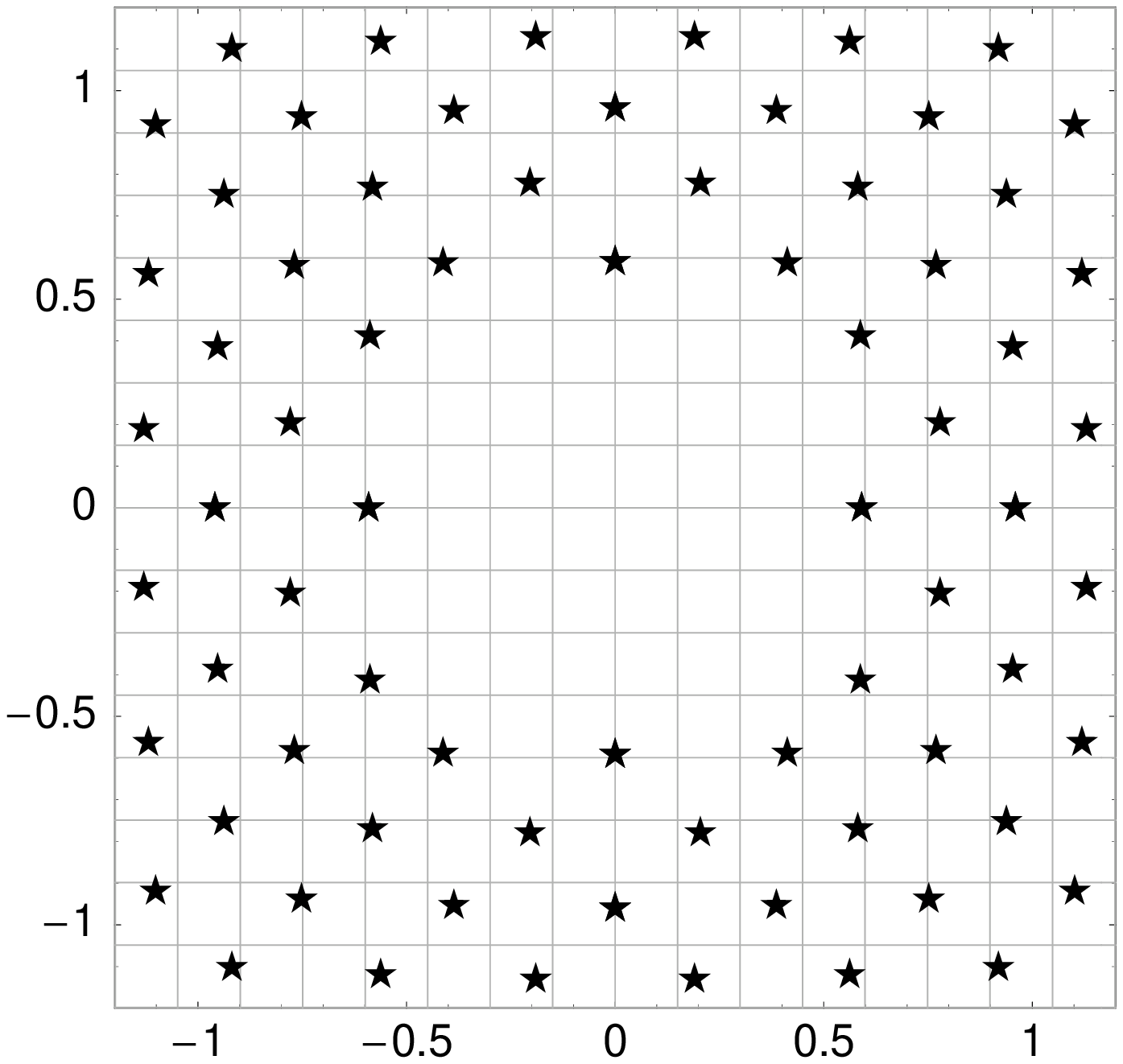}}
\put(48,0){$\RRe[q_3^{1/4}]$}
\put(18,27){\rotatebox{90}{$\IIm[q_3^{1/4}]$}}
\put(113,0){$\RRe[q_4^{1/2}]$}
\put(83,27){\rotatebox{90}{$\IIm[q_4^{1/2}]$}}
\end{picture}
}
\caption[The spectra of the conformal charges 
for $N=4$ with $\theta_4=\pi$ 
$\ell_3=0$ and $\ell_4=1$.]{The spectra of 
the conformal charges for $N=4$ with $\theta_4=\pi$,
$\ell_3=0$ and $\ell_4=1$. 
On the left panel the spectrum of $q_4^{1/4}$, while on the right
panel the spectrum of $q_3^{1/2}$}
\lab{fig:l30k21}
\end{figure}

\medskip
\begin{figure}[ht!]
\centerline{
\begin{picture}(165,60)
\put(21,2){\epsfysize6.0cm \epsfbox{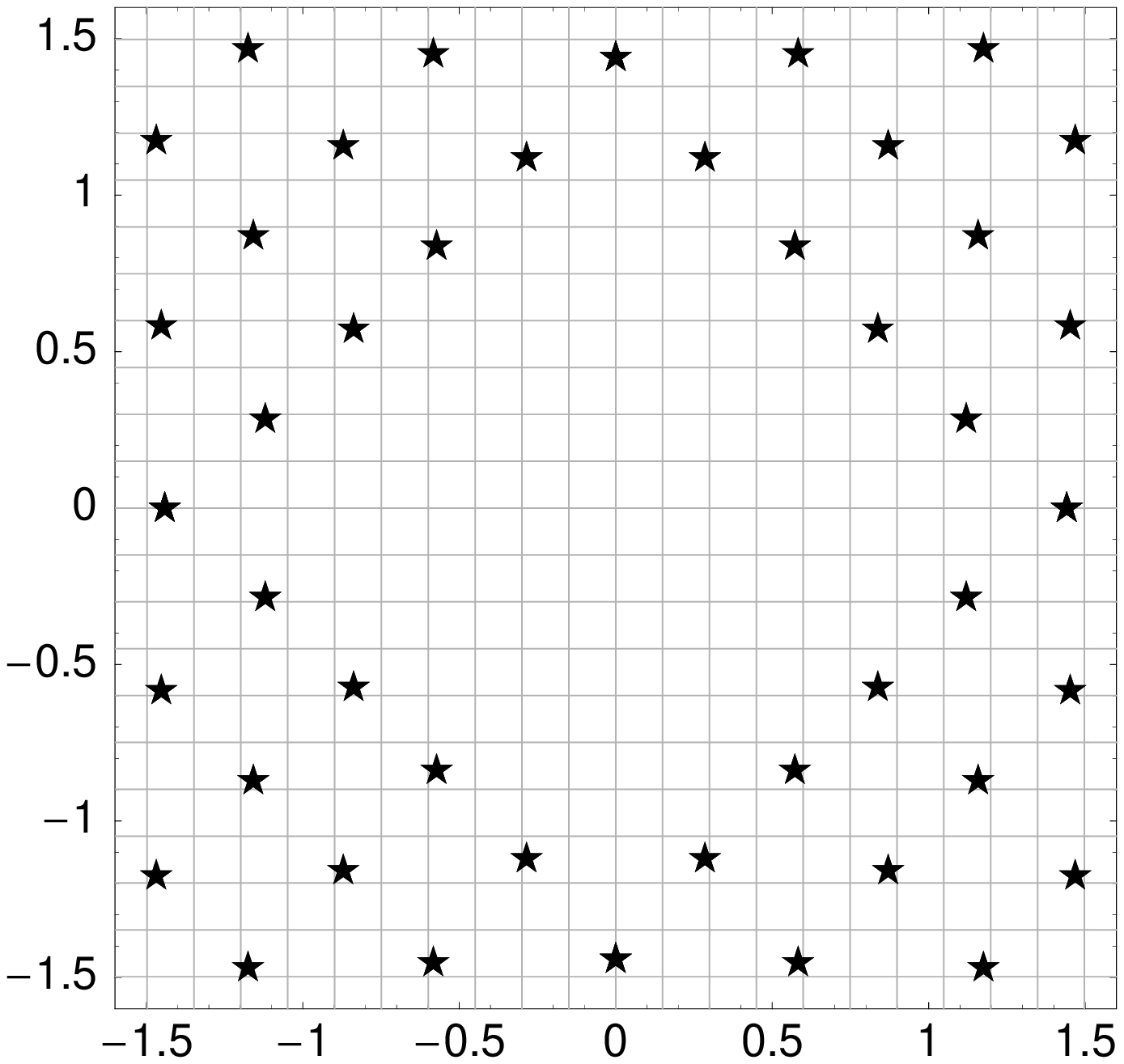}}
\put(85,2){\epsfysize6.1cm \epsfbox{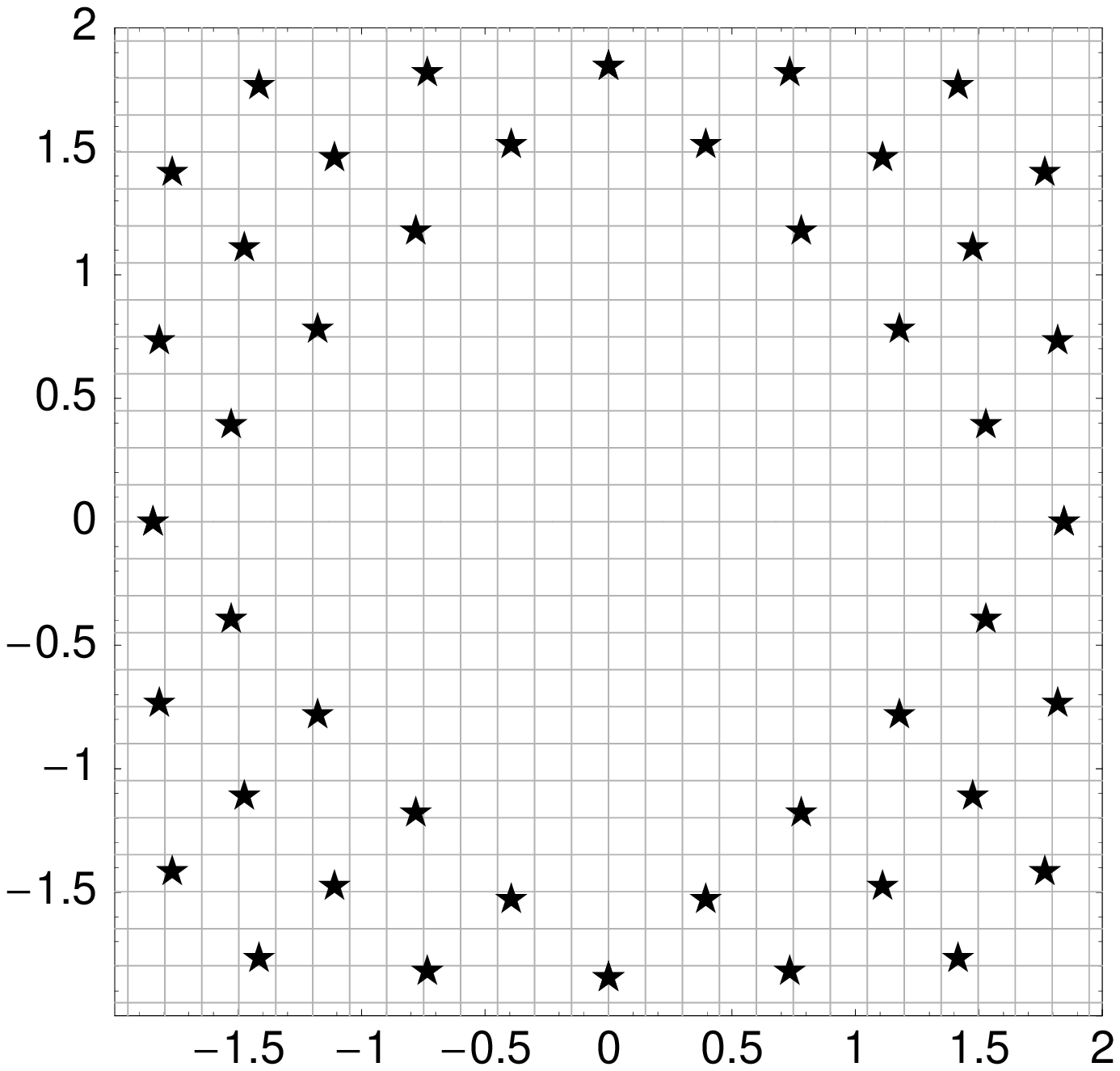}}
\put(48,0){$\RRe[q_3^{1/4}]$}
\put(18,27){\rotatebox{90}{$\IIm[q_3^{1/4}]$}}
\put(113,0){$\RRe[q_4^{1/2}]$}
\put(83,27){\rotatebox{90}{$\IIm[q_4^{1/2}]$}}
\end{picture}
}
\caption[The spectra of the conformal charges 
for $N=4$ with $\theta_4=\pi$, 
$\ell_3=0$ and $\ell_4=2$.]{The spectra of 
the conformal charges for $N=4$ with $\theta_4=\pi$,
$\ell_3=0$ and $\ell_4=2$.  
On the left panel the spectrum of $q_4^{1/4}$, while on the right
panel the spectrum of $q_3^{1/2}$}
\lab{fig:l30k22}
\end{figure}

\subsection{Winding lattices with $\ell_3 \ne 0$}

In the case with $\ell_3\ne0$ we have much more complicated 
situation than for $\ell_3=0$.
According to (\ref{eq:q3-rphi}) the angles $\phi_3$ and $\phi_4$
defined in (\ref{eq:qrphi}) are no more equal. Moreover,
they start to depend on the scale $\lambda=r_3/r_4$.

\medskip
\begin{figure}[ht!]
\centerline{
\begin{picture}(165,60)
\put(23,4){\epsfysize5.7cm \epsfbox{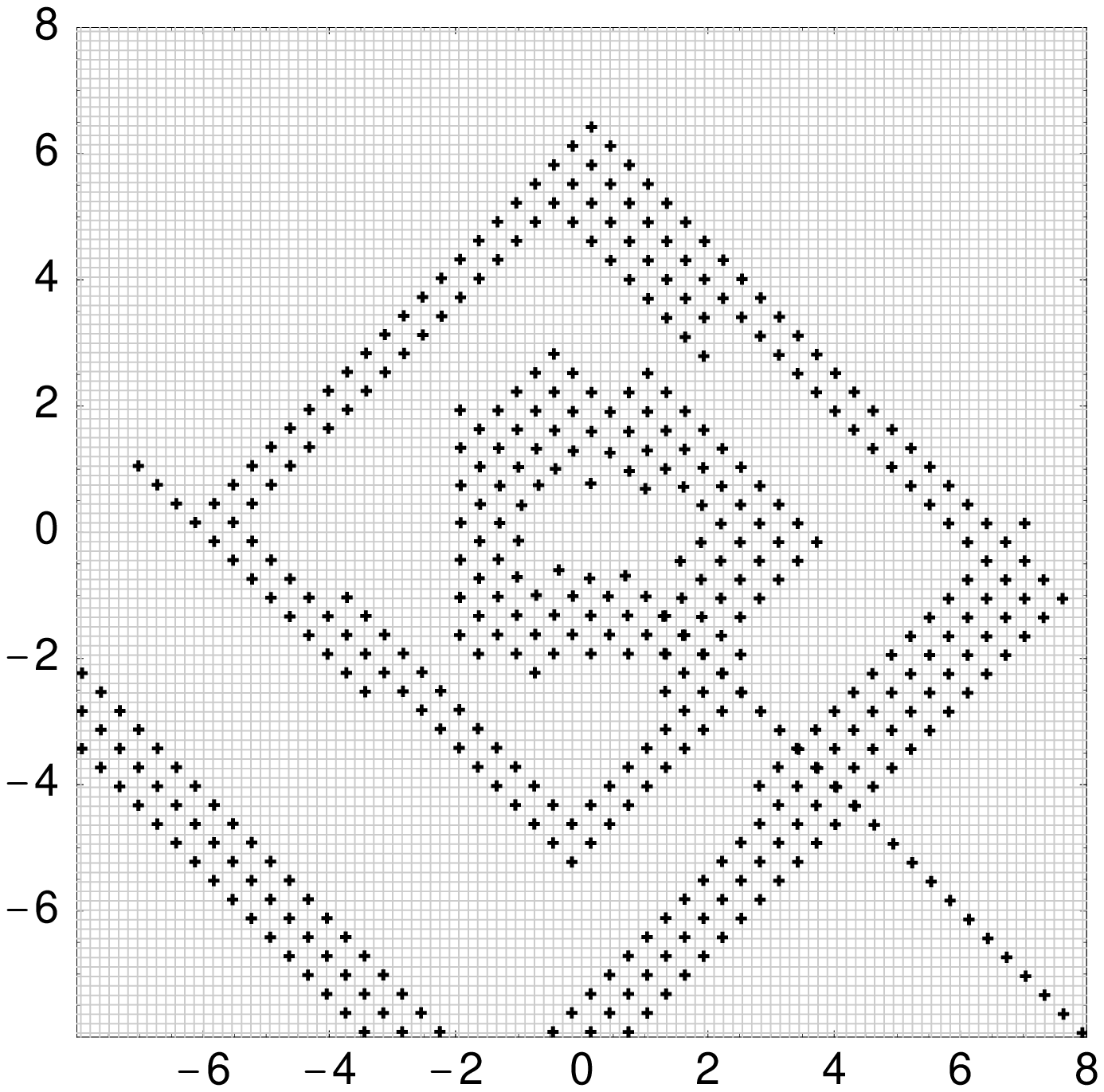}}
\put(85,2){\epsfysize6.0cm \epsfbox{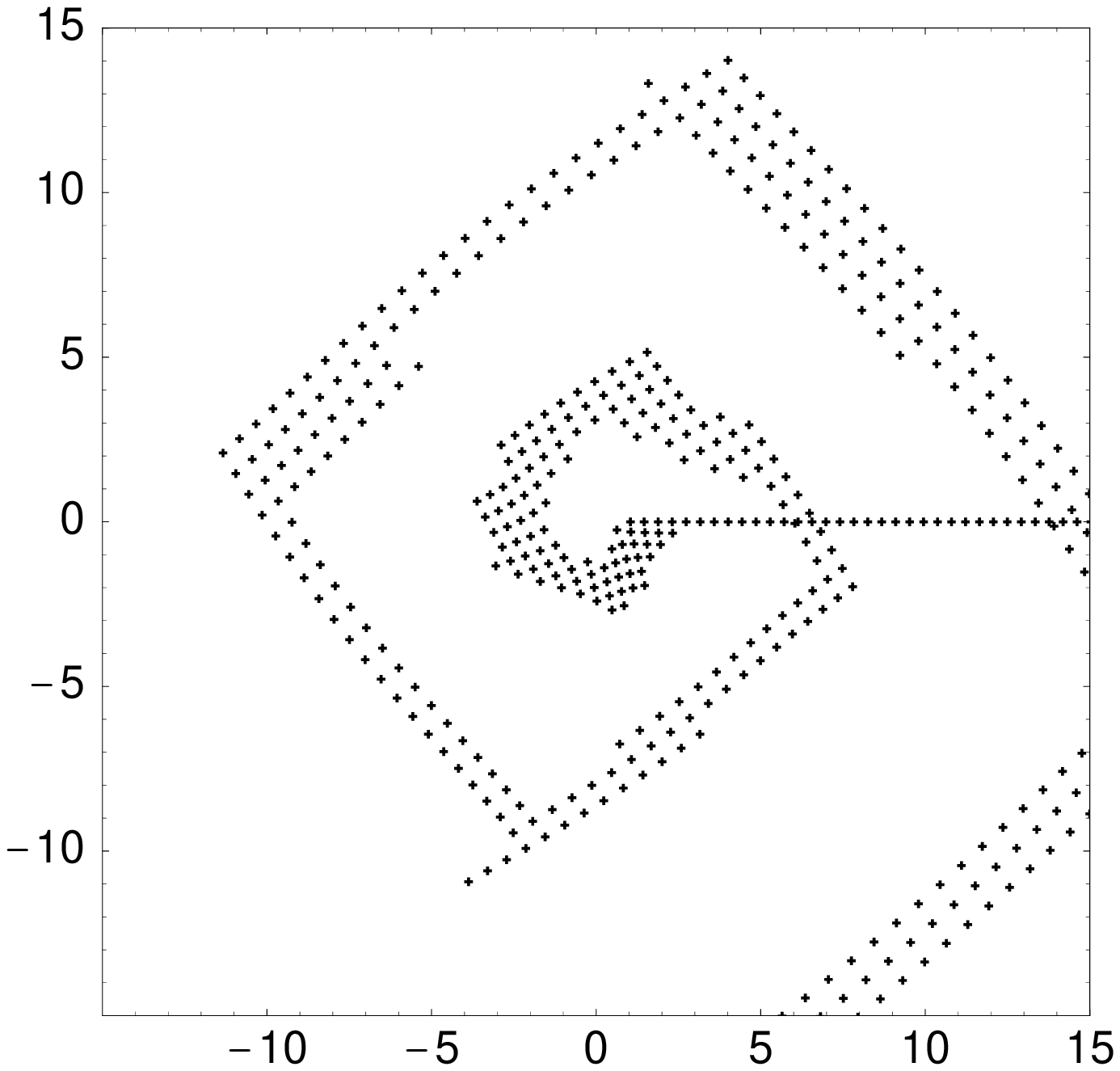}}
\put(48,0){$\RRe[q_3^{1/4}]$}
\put(18,27){\rotatebox{90}{$\IIm[q_3^{1/4}]$}}
\put(113,0){$\RRe[q_4^{1/2}]$}
\put(82,27){\rotatebox{90}{$\IIm[q_4^{1/2}]$}}
\end{picture}
}
\caption[The winding spectrum of the conformal charges 
for $N=4$ with $h=1/2$, $\theta_4=- \pi/2$ and 
$\ell_3=1$.]{The winding 
spectrum of the conformal charges for $N=4$ and $h=1/2$ 
with $\theta_4=-\pi/2$ and 
$\ell_3=1$. On the left panel the spectrum of $q_4^{1/4}$, while on the right
panel the spectrum of $q_3^{1/2}$}
\lab{fig:l31k31}
\end{figure}

An example of such a lattice, with $\ell_3=1$ and $\theta_4=-\pi/2$
we show in Figure \ref{fig:l31k31}.
For this case,
the $q_4^{1/4}-$lattice is defined by (\ref{eq:q4-quan})
with $\ell_1$ and $\ell_2$ odd integer numbers satisfying
$\ell_1+\ell_2 \in 4 \mathbb{Z}$.
In Figure \ref{fig:l31k31}, in order to present the correspondence 
between the $q_3^{1/2}-$ and $q_4^{1/4}-$lattice, we depict only some vertices
of the lattice which extends in the whole plane of the conformal charges
except the place nearby the origin $q_4=q_3=0$.
As we can see the $q_3^{1/2}-$lattice is still a square-like one
but it winds around the origin $q_3^{1/2}=0$.
Looking at Figure \ref{fig:l31k31}, let us start from
$\phi_3=0$ and $\phi_4=-\pi/4$ where $\phi_3$ and $\phi_4$ are defined
in (\ref{eq:qrphi}). 
Thus, the difference $\phi_3-\phi_4=\pi/4$
so that our scale at the beginning is $\lambda=\sqrt{2}$. 
In this region the vertices of the $q_3^{1/2}-$lattice 
are in the nearest place to the origin.
When we go clockwise 
around the origin of the lattices by decreasing $\phi_4$,
we notice that $\phi_3$ also decreases but much slower.
Thus, according to (\ref{eq:q3-rphi}) 
the difference $\phi_3-\phi_4$ changes. Moreover,
due to (\ref{eq:q3-rphi}), the scale $\lambda$ also continuously rises.
Therefore, the $q_3^{1/2}-$lattice winds in a different way 
than the $q_4^{1/4}-$lattice. After one revolution the vertices
of the $q_4^{1/4}-$lattices are at similar places as those 
with $\phi_4$ decreased by $2 \pi$. This provides
additional spurious degeneration in $q_4^{1/4}$. However, 
after revolution by the angle $2 \pi$, the vertices in $q_3^{1/2}-$space 
have completely different conformal charges $q_3$. The spurious degeneration
in the $q_3^{1/2}-$lattice does not appear.

We have to add that we obtain the second symmetric structure 
when we go in the opposite direction, \ie anti-clockwise.
Moreover,
the winding lattice, like all other lattices,
extends to infinity on the $q_3^{1/2}-$
and $q_4^{1/4}-$plane and  does not have vertices 
in vicinity of the origin, $q_3=q_4=0$. However, 
the radii of these empty spaces grow with 
$\phi_{3,4}$.

Additionally, due to symmetry of the spectrum (\ref{eq:qkcsym})
we have a twin lattice with $q_k\rightarrow q_k^{\ast}$. 
Furthermore,
the second symmetry (\ref{eq:qkmsym}) produces another lattice
rotated in the $q_3^{1/2}-$space by an angle $\pi/2$. 
Notice that this symmetry (\ref{eq:qkmsym}) exchanges quasimomentum
${\theta_4 \leftrightarrow 2\pi - \theta_4 ({\rm mod}~ 2\pi)}$. 
Thus, the spectrum for $\theta_4=\pi/2$
is congruent with the spectrum of $\theta_4=-\pi/2$ but it is rotated
in the $q_3^{1/2}-$space by $\pi/2$.

\medskip
\begin{figure}[ht!]
\centerline{
\begin{picture}(165,60)
\put(23,4){\epsfysize5.7cm \epsfbox{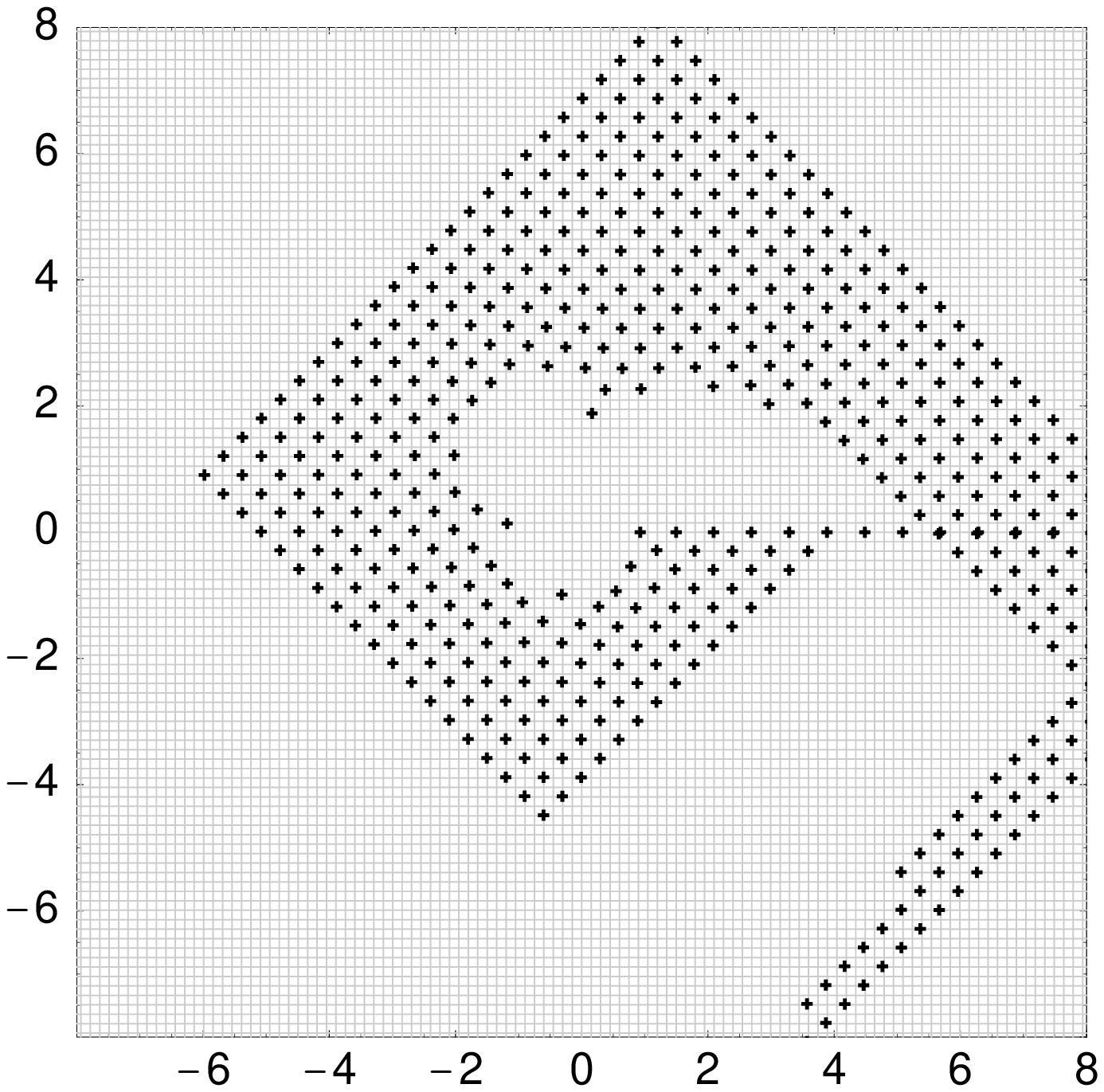}}
\put(85,2){\epsfysize6.0cm \epsfbox{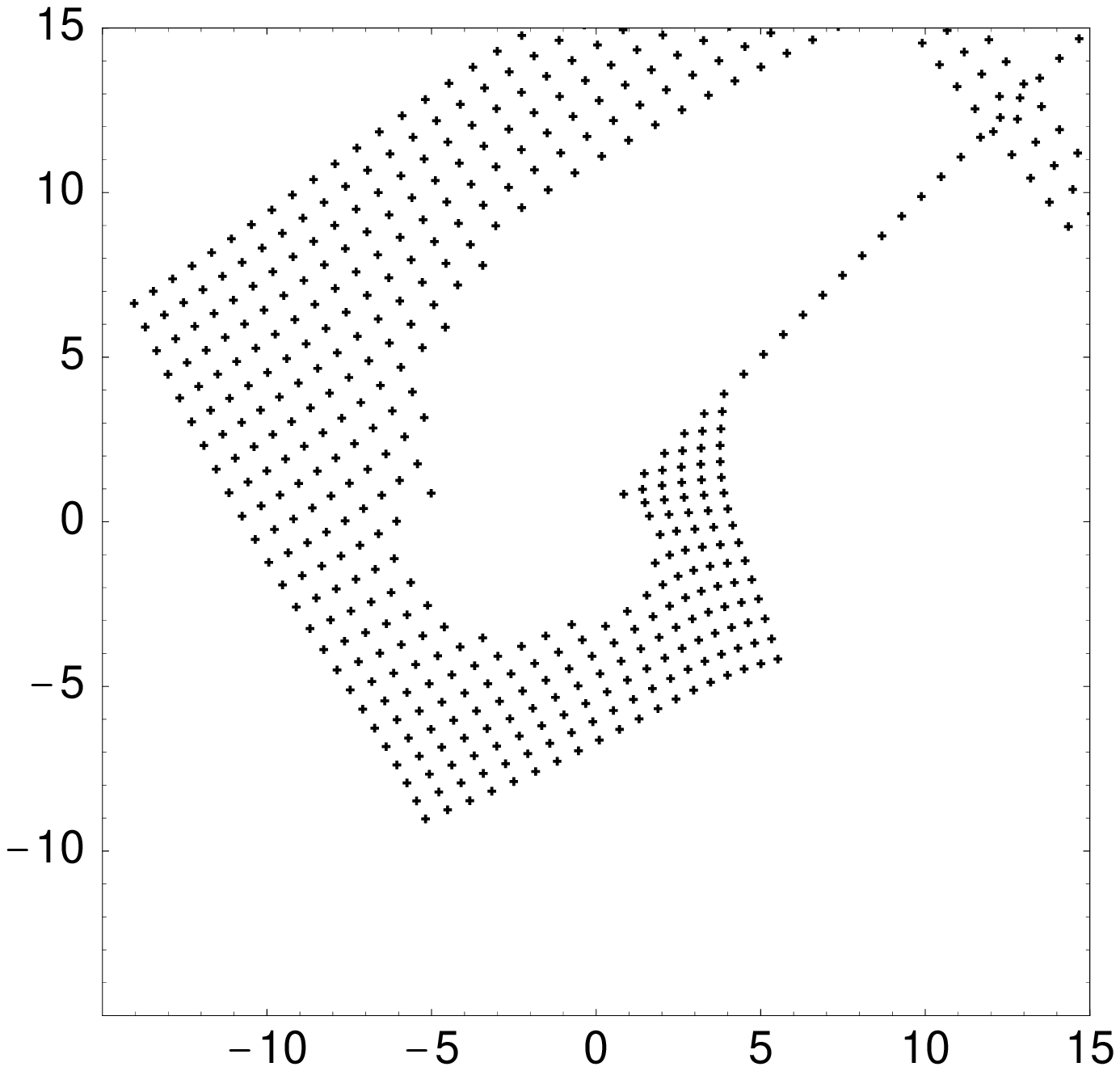}}
\put(48,0){$\RRe[q_3^{1/4}]$}
\put(18,27){\rotatebox{90}{$\IIm[q_3^{1/4}]$}}
\put(113,0){$\RRe[q_4^{1/2}]$}
\put(82,27){\rotatebox{90}{$\IIm[q_4^{1/2}]$}}
\end{picture}
}
\caption[The winding spectrum of the conformal charges 
for $N=4$ and $h=1/2$ with $\theta_4=0$ and 
$\ell_3=2$.]{The winding 
spectrum of the conformal charges for $N=4$ with $h=1/2$, $\theta_4=0$ and 
$\ell_3=2$. On the left panel the spectrum of $q_4^{1/4}$ while on the right
panel the spectrum of $q_3^{1/2}$}
\lab{fig:l32k0}
\end{figure}

As we said before 
for $\theta_4 \pm \pi/2$ we have winding spectra with odd $\ell_3$.
Similarly,
for even $\ell_3 \ne 0$ we have winding spectra with $\theta_4=0,\pi$,
thus, the lattice with the lowest non-zero $\ell_3$ corresponds to
$|\ell_3|=2$.
We present some points of this spectrum in Fig.\ \ref{fig:l32k0},
for which $\ell_1$, $\ell_2$ in 
(\ref{eq:q4-quan})
are even and 
$\ell_1+\ell_2 \in 4 \mathbb{Z}+2$.
In this case we start with $\phi_3=\pi/4$ and $\phi_4=0$ which implies
the beginning scale $\lambda=\sqrt{2}$. 
The spectra wind as in the previous case.

\begin{figure}[ht!]
\centerline{
\begin{picture}(165,60)
\put(23,4){\epsfysize5.7cm \epsfbox{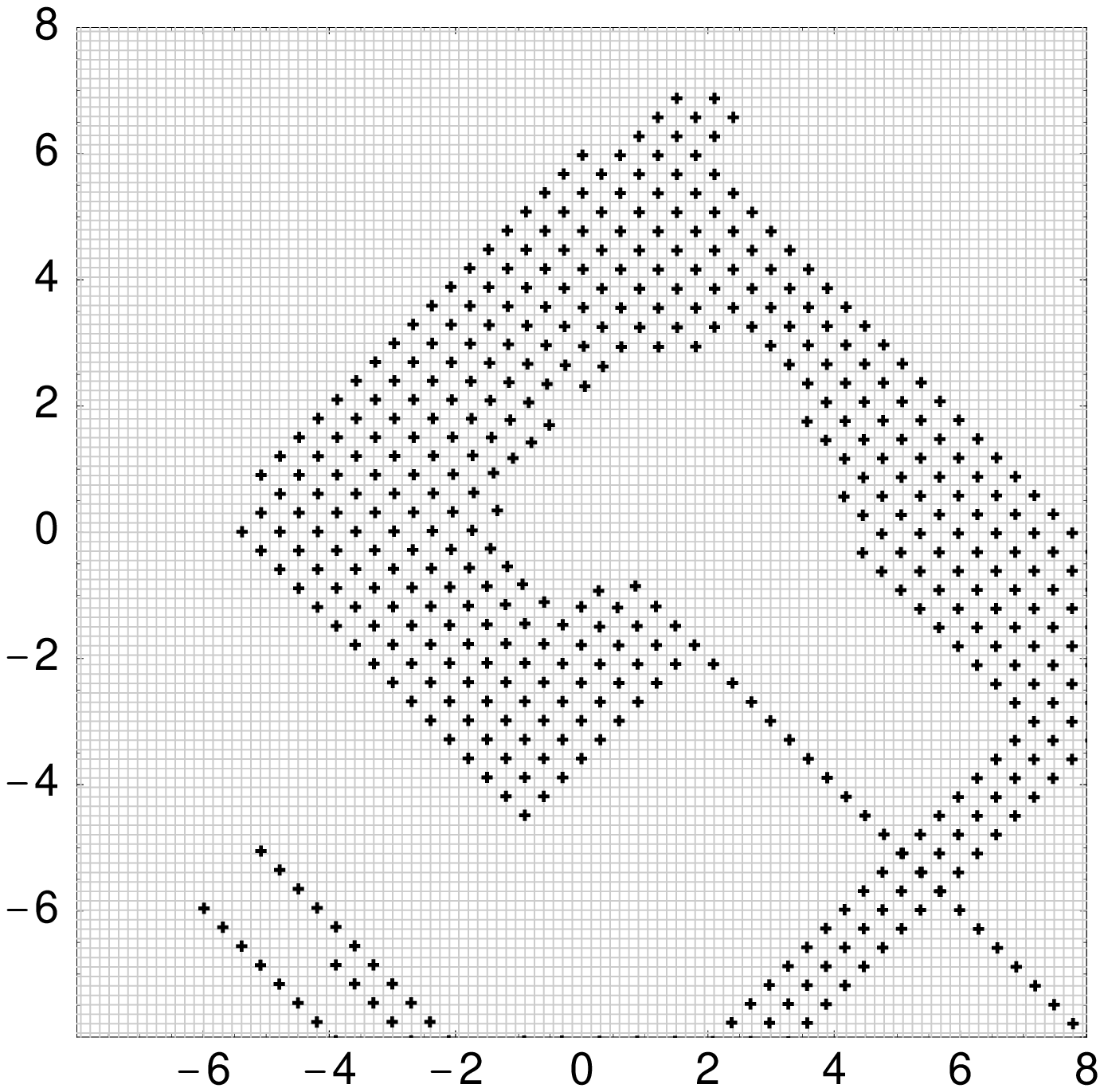}}
\put(85,2){\epsfysize6.0cm \epsfbox{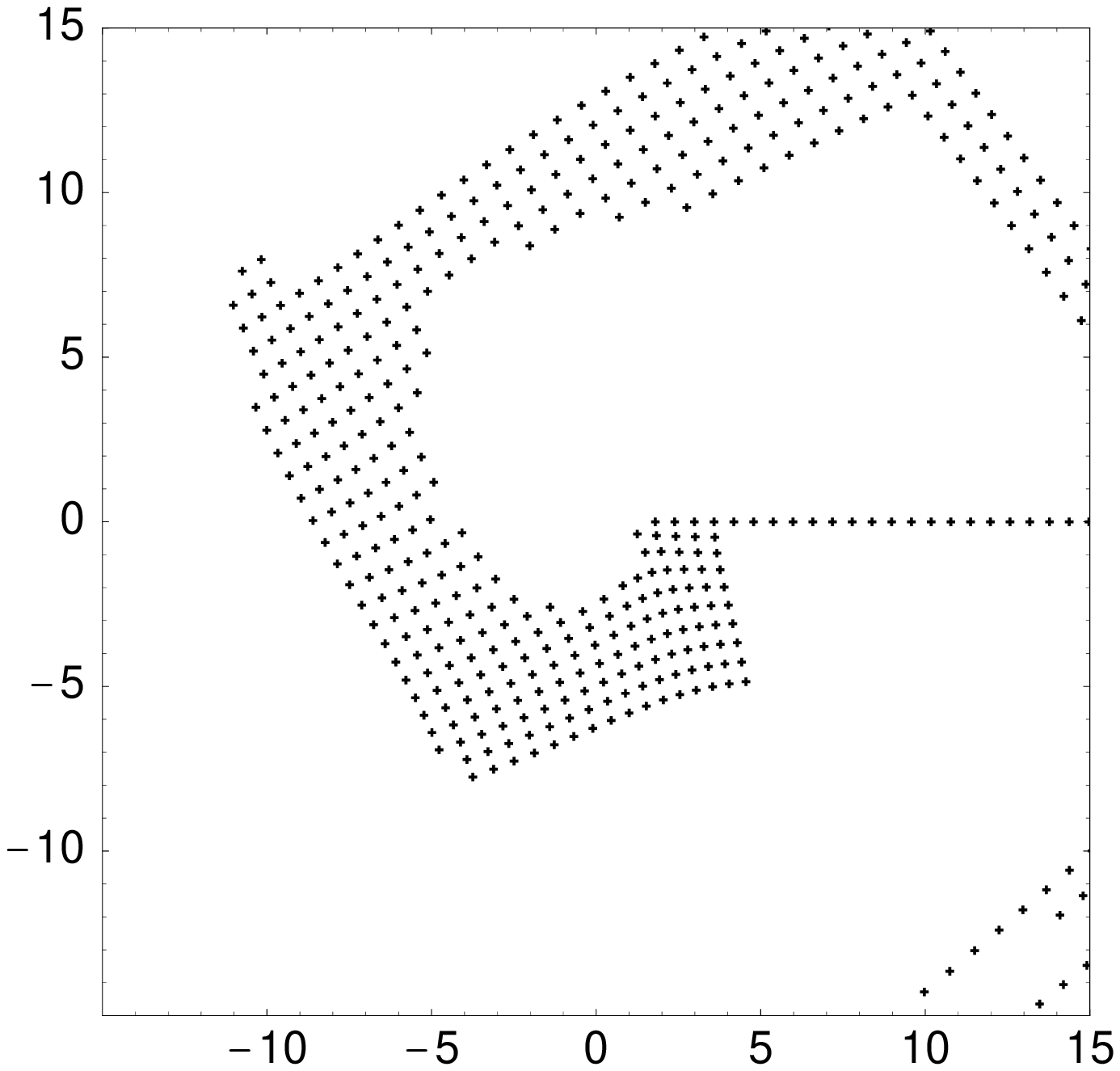}}
\put(48,0){$\RRe[q_3^{1/4}]$}
\put(18,27){\rotatebox{90}{$\IIm[q_3^{1/4}]$}}
\put(113,0){$\RRe[q_4^{1/2}]$}
\put(82,27){\rotatebox{90}{$\IIm[q_4^{1/2}]$}}
\end{picture}
}
\caption[The winding spectrum of the conformal charges 
for $N=4$ and $h=1/2$ with $\theta_4=\pi$ and 
$\ell_3=2$.]{The winding 
spectrum of the conformal charges for $N=4$
with $h=1/2$, $\theta_4=\pi$ and 
$\ell_3=2$. On the left panel the spectrum of $q_4^{1/4}$ while on the right
panel the spectrum of $q_3^{1/2}$}
\lab{fig:l32k2}
\end{figure}

Similarly,
for $\theta_4=\pi$ we have also spectrum with the lowest  $|\ell_3|=2$.
It is defined by (\ref{eq:q4-quan}) with $\ell_1$, $\ell_2$ even
and $\ell_1+\ell_2 \in 4 \mathbb{Z}$.
In this case the angles $\phi_3=0$ and $\phi_4=-\pi/4$ so we also have
the beginning scale $\lambda=\sqrt{2}$, as depicted in Figure \ref{fig:l32k2}.
In order to describe the winding spectra better we may introduce
an integer parameter $\ell_4$ which helps us to number the 
overlapping winding planes
of the spectra and name spuriously-degenerated vertices
in the $q_4^{1/4}-$plane.

To sum up, even for a given quasimomentum we have many lattices
which overlap, so that vertices of the lattices,
especially in $q_3^{1/2}-$space, make an impression
of being randomly distributed.
However, as we have shown
above, those spectra may be distinguished and finally described
by (\ref{eq:q4-quan}) and (\ref{eq:q3-quan}).
Still, there is a lack of one nontrivial WKB condition which
would uniquely explain the structure
of the resemblant and winding lattices.

\subsection{Corrections to WKB}

Let us consider the spectrum of 
the conformal charge $q_4$ for $N=4$ with $q_3=0$ and ${h=\frac{1+n_h}{2}}$.
It turns out that for $n_h \ne 0$ it has similar square-like
lattice structure like that with $n_h=0$,
see Fig.\ \ref{fig:WKB-N4} on the right panel.
Similarly to the case with three reggeized gluons
\ci{Kotanski:2006ec} 
we have evaluated the conformal charges $q_4$
with even $n_h$ with high precision.
We have done it separately for $q_4$ with $\IIm[q_4]=0$
and $\RRe[q_4]=0$.
Next, we have fitted
coefficient expansion of the WKB series, $a^{(r)}_k$ and $a^{(i)}_k$, 
respectively.

In \ci{Derkachov:2002pb} the series formula for $q_4^{1/4}$
looks as follows
\begin{equation}
q_{4}^{1/4}=\frac{\pi^{3/2}}{2 \Gamma ^{2}(1/4)}
{\cal Q}(\mybf n)\left[1+\frac{b}{\left|{\cal Q}(\mybf n)\right|^{2}}+
\sum _{k=2}^{\infty }{a_{k}
\left(\frac{b}{\left|{\cal Q}(\mybf n)\right|^{2}}\right)^{k}}\right]\,,
\lab{eq:korq4}
\end{equation}
where 
\begin{equation}
{\cal Q}(\mybf n)
=\sum _{k=1}^{4}{n_{k}e^{i\pi (2k-1)/4}}
=\left( \frac{\ell_1}{\sqrt{2}} + i \frac{\ell_2}{\sqrt{2}}\right)
\lab{eq:Qn4}
\end{equation}
and $\ell_1$,$\ell_2$, ${\mybf n}=\{n_1,\ldots,n_N\}$ are integer.

\begin{table}[ht]
\begin{center}
$\begin{array}{|c|c||r|r|r|r|r|} 
    \hline
n_h & \mbox{coef.} & k=2 \qquad & k=3 \qquad & k=4 \quad & k=5\quad
 \\ \hline \hline
0
&a^{(r)}_k           &     2.9910566246  & -24.021689 &   91.591 &  645.5 \\
&a^{(i)}_k           &    -4.9910566246  &  28.021689 & -148.830 & 1656.7 \\
&a^{(r)}_k+a^{(i)}_k &    -2.0000000000  &   4.000000 &  -57.239 & 2302.2 \\
    \hline
2
&a^{(r)}_k           &     -1.3991056625 &   4.674008 &   -4.516 &   -95.7 \\
&a^{(i)}_k           &     -0.6008943375 &  -0.674008 &   20.388 &  -200.8 \\
&a^{(r)}_k+a^{(i)}_k &     -2.0000000000 &   4.000000 &   15.872 &  -296.5 \\
    \hline
4
&a^{(r)}_k           &      -1.351212983 &   3.462323 &  -11.322 &  43.164 \\ 
&a^{(i)}_k           &      -0.648787017 &   0.537677 &   -0.096 &   0.611 \\
&a^{(r)}_k+a^{(i)}_k &      -2.000000000 &   4.000000 &  -11.418 &  43.775 \\  
    \hline
6
&a^{(r)}_k           &     -0.8882504145 & 1.883461 &  -5.4248 & 18.081 \\
&a^{(i)}_k           &     -1.1117495855 & 2.116539 &  -4.8117 & 12.091 \\
&a^{(r)}_k+a^{(i)}_k &     -2.0000000000 & 4.000000 & -10.2365 & 30.172 \\
    \hline
8
&a^{(r)}_k           &     -0.6326570719 &  1.197694 & -2.94579 &   8.3372 \\
&a^{(i)}_k           &     -1.3673429281 &  2.802305 & -5.96129 &  12.1238 \\
&a^{(r)}_k+a^{(i)}_k &     -2.0000000000 &  4.000000 & -8.90708 &  20.4610 \\
    \hline
\end{array}$
\end{center}
\caption[The fitted coefficient to the series formula of $q_4^{1/4}$]
{The fitted coefficient to the series formula of $q_4^{1/4}$ 
(\ref{eq:korq4}) with  $n_h=0,2,4,6$ and $8$}
\lab{tab:q4coefs}
\end{table}

Here for $N=4$, 
similarly to the $N=3$ case 
in Ref.~\ci{Kotanski:2006ec},
we have a different expansion parameter 
$b/\left|{\cal Q}(\mybf n)\right|^{2}$
and
in this case the parameter
\begin{equation}
b= \frac{4}{\pi}\left({q_{2}}^{\ast}-\frac{5}{4}\right)
\lab{eq:coef4}
\end{equation}
is decreased by $5/4$.
The expansion coefficients of the series (\ref{eq:korq4}) 
are shown in Table \ref{tab:q4coefs}.
Similarly to \ci{Derkachov:2002pb} 
the coefficients $a_0=1$ and $a_1=1$. 
The remaining coefficients 
depend on $n_h$. 
Moreover, the coefficients $a_k$ with $k>1$ are different for real
$q_4^{1/4}$ and for imaginary $q_4^{1/4}$
but one may notice that 
for $k=1,2$ the sum $a_k^{(r)}+a_k^{(i)} \in \mathbb{Z}$
and does not depend on $n_h$.
Thus, to describe 
the quantized values of $q_4^{1/4}$ 
more generally one has
to use  both sets of the coefficients, $a_k^{(r)}$ and $a_k^{(i)}$,
or perform the expansion with two small independent parameters, \ie
$q^{\ast}_2/|{\cal Q}({\mybf n})|^2$ and $1/|{\cal Q}({\mybf n})|^2$.

Using the series (\ref{eq:korq4}) with (\ref{eq:coef4}) and coefficients from
Table \ref{tab:q4coefs} gives good approximation of the conformal charges 
$q_4$ with $q_3=0$. However, in order to have a better precision
one has to introduce an additional expansion parameter.

\section{Summary and Conclusions}

In this paper we discussed four Reggeon states which appear
is scattering amplitude of strongly interacting particles
in high energy Regge limit of QCD (\ref{eq:Rlim}),
\ie in Generalized Leading Logarithm Approximation (GLLA)
\ci{Bartels:1980pe,Kwiecinski:1980wb,Jaroszewicz:1980mq}.
Due to the colour factor
description of $N-$Reggeon system is a complicated task especially for
$N>3$. To this end multi-colour limit is performed.
In this limit the Reggeon Hamiltonian 
separates into two one-dimensional equations and as a result 
the multi-Reggeon system get completely solvable. 
However, in order to solve it one has to use advanced
integrable-model methods.

In the multi-colour limit the equation for  the $N-$Reggeon wave-function
takes a form of Schr\"odinger equation (\ref{eq:Schr}) for
the non-compact XXX Heisenberg magnet model
of $\SL(2,\mathbb{C})$ spins $s$
\ci{Takhtajan:1979iv,Faddeev:1979gh,KBI}. 
Its Hamiltonian describes the nearest neighbour interaction
of the Reggeons \ci{Lipatov:1993yb,Faddeev:1994zg} 
propagating in the two-dimensional transverse-coordinates space 
(\ref{eq:trcoords}).
The system has 
a hidden cyclic and mirror permutation symmetry (\ref{eq:PMsym}). 
It also possesses
the set of the $(N-1)$ of integrals of motion, which 
are eigenvalues of conformal charges  \ci{Derkachov:2001yn},
$\oq{k}$ and $\oqb{k}$,
(\ref{eq:qks0})--(\ref{eq:qks1}). Therefore, the operators of
conformal charges commute with each other and with
the Hamiltonian and they possess a common set of
the eigenstates.

To solve the Reggeon problem for more than three particles
one uses the more sophisticated
technique, \ie the Baxter $Q-$operator method 
\ci{Derkachov:2001yn}.
It relies on the existence of the operator $\mathbb{Q}(u,\wbar{u})$
depending on the pair of complex spectral parameters 
$u$ and $\wbar u$. The Baxter $Q-$operator 
has to commute with itself (\ref{eq:comQQ}) and with 
the conformal charges (\ref{eq:comQt}). It also has to satisfy
the Baxter equations
(\ref{eq:Baxeq})--(\ref{eq:Baxbeq}).
Furthermore, the $Q-$operator has
well defined analytical properties, \ie known pole structure 
(\ref{eq:upoles}) 
and asymptotic behaviour at infinity (\ref{eq:Qanalb}).
The above conditions fix the $Q-$operator
uniquely and allow to quantize the integrals $q_k$.
In turns out \ci{Derkachov:2001yn} 
that the Reggeon Hamiltonian 
can be rewritten in terms of 
Baxter $Q-$operator (\ref{eq:HNQQ}).
Therefore, combining together the solutions 
of the Baxter equations and the conditions for $q_k$
with the Schr\"odinger equation, we can calculate the 
energy spectrum (\ref{eq:energy}). Moreover,
we are able to determine the quasimomentum of the
eigenstates (\ref{eq:quasQ}), \ie
the observable which defines the properties of the
state with respect 
to the cyclic permutation symmetry (\ref{eq:PMsym}).

In order to obtain the exact values of $q_k$
we have used the method \ci{Derkachov:2002pb,Derkachov:2002wz}
which consists in rewriting the
Baxter equations and the other conditions for 
$Q-$operator eigenvalues as the $N-$order
differential equation (\ref{eq:Eq-1}).
Solving this equation one obtains the conditions 
for quantized $q_k$ (\ref{eq:C1-C0}) and 
the formulae for the energy (\ref{eq:E-fin}) 
and the quasimomentum (\ref{eq:parity-qc}).

In the case with $N=4$ particles we have constructed
the spectrum for $n_h=0$
depicting complicated interplay between
the conformal charges: $q_3$ and $q_4$.
Such a complete analysis has been performed here for the first time. 
Earlier, for $N=4$ full spectrum of $q_4$ was shown in 
Ref. \ci{Derkachov:2002wz}
(however the corresponding $q_3$ spectrum was not discussed)
and some values of $q_4$ were found in Ref. \ci{DeVega:2001pu}.
The spectrum of $q_4^{1/4}$ has a structure of square-like lattice 
(\ref{eq:WKB-N4}), Fig.\ \ref{fig:Q4}.
In this case the spectrum of $q_3$ is more complicated. It turns
out that it has a few possible forms. 
Firstly, there are simple states with $q_3=0$, Fig.\ \ref{fig:WKB-N4}.
Moreover, we have found 
the $q_3^{1/2}-$lattices whose distribution of vertices
is similar to the distribution of vertices of 
$q_4^{1/4}-$lattice. We have called this structures as resemblant lattices,
Figs.~\ref{fig:l30k01}-\ref{fig:l30k22}.
Secondly we have demonstrated the existence of $q_3^{1/2}-$lattices 
that may be called winding lattices.
They wind around the origin and in course of this winding the distance
between vertices increases and also the lattice goes away from the origin,
Figs.\ \ref{fig:l31k31}-\ref{fig:l32k2}.
The WKB approximation does not describe these structures (\ref{eq:q3-quan})
exactly because
one quantization condition is missing.
Finally, we have also considered the descendent states 
with $q_4=0$, Fig.\ \ref{fig:WKB-N4}.
They have quasimomentum $\theta_4=\pi$ and the structure
of their $q_3^{1/3}-$lattices is the same as for
$q_3^{1/3}-$lattices 
in the three-Reggeon case with $\theta_4=0$ \ci{Kotanski:2005ci,Kotanski:2006ec}. 
For the $N=4$ Reggeon states with $q_3=0$
we have calculated corrections to the WKB approximation 
(\ref{eq:korq4}), Table \ref{tab:q4coefs}.
Similarly to the $N=3$ case \ci{Kotanski:2006ec} they come from the fact 
of using only one expansion parameter $\eta$ 
for two different conformal charges, $q_4$ and $q_2$.
One has to notice that these corrections are subleading
in comparison to the WKB limit \ci{Derkachov:2002pb}.


The $Q-$Baxter method
is complicated. However,
it allows us to solve the reggeized gluon state problem
for more than $N=3$ particles. The above calculations
are of interest not only for perturbative QCD
but also to statistical physics 
as the $\SL(2,\mathbb{C})$ non-compact XXX Heisenberg spin magnet model
\ci{Takhtajan:1979iv,Faddeev:1979gh,KBI}.

In order to find the full
scattering amplitude one has to calculate the impact factors, \ie
overlaps between the Reggeon wave function and the wave functions
of the scattered particles.
These impact factors strongly depend on the scattered system
and they are hard to calculate due to a large number of 
integrations. Thus, this work can be a first step in computing
of contribution of the four-Reggeon state 
to the hadron scattering amplitude.

\section*{Acknowledgements}
I would like to warmly thank to
Micha{\l} Prasza{\l}owicz
for fruitful discussions and help during writing this work.
I am very grateful to G.P. Korchemsky,
A.N.Manashov and S.\'E. Derkachov
whom I worked in an early state of this project.
I also thank to 
Jacek Wosiek
for illuminating discussions.
This work was supported by
KBN PB 2-P03B-43-24,
KBN PB 0349-P03-2004-27 and
KBN PB P03B-024-27:2004-2007.


\end{document}